\documentclass[aps,prf,twocolumn,superscriptaddress]{revtex4-2}
\usepackage{amssymb,amsmath,esint}
\usepackage{graphicx}
\usepackage[version=3]{mhchem}
\usepackage[ulem=normalem]{changes}

\begin{document}
\graphicspath{{}}

\title{Triple line destabilization -- Tuning film thickness through meniscus curvature}

\author{P. Hayoun}
\affiliation{Soft Matter Sciences and Engineering, ESPCI Paris, PSL University,
Sorbonne Universit\'{e}, CNRS, F-75005 Paris, France}
\affiliation{Saint-Gobain Research Paris, 39 quai Lucien Lefranc, 93330 Aubervilliers, France}
\author{A. Letailleur}
\affiliation{Saint-Gobain Research Paris, 39 quai Lucien Lefranc, 93330 Aubervilliers, France}
\author{J. Teisseire}
\affiliation{Saint-Gobain Research Paris, 39 quai Lucien Lefranc, 93330 Aubervilliers, France}
\affiliation{Surface du Verre et Interfaces, CNRS/Saint-Gobain, 39 quai Lucien Lefranc, 93330 Aubervilliers, France}
\author{F. Lequeux}
\affiliation{Soft Matter Sciences and Engineering, ESPCI Paris, PSL University,
Sorbonne Universit\'{e}, CNRS, F-75005 Paris, France}
\author{E. Verneuil}
\affiliation{Soft Matter Sciences and Engineering, ESPCI Paris, PSL University,
Sorbonne Universit\'{e}, CNRS, F-75005 Paris, France}
\author{E. Barthel}
\affiliation{Soft Matter Sciences and Engineering, ESPCI Paris, PSL University,
Sorbonne Universit\'{e}, CNRS, F-75005 Paris, France}


\date{\today}

\begin{abstract}
For partially wetting fluids, previous results suggest that the thickness and the dewetting velocity of gravity driven films are uniquely determined by the triple line dynamics. In contrast, when flushing aqueous liquids through polymer tubes, our measurements show that the dewetting velocity and thickness can be selected. The control parameter is pressure, \emph{i.e.} the macroscopic curvature of the meniscus. Our results demonstrate directly the major role played by the macroscopic geometry in the stability of a dynamic meniscus as predicted by Eggers (Phys. Rev. Lett. 93, 094502 (2004)).
\end{abstract}


\maketitle


\section{Introduction}
When a solid substrate is immersed in a liquid bath, capillary forces distort the liquid/air interface close to the solid and a static meniscus is established. The characteristic length which emerges from the competition between capillary and hydrostatic pressures is the capillary length $l_c=(\gamma/{\rho g})^{1/2}$ where $\gamma$ is liquid surface tension, $\rho$ the density and $g$ the acceleration of gravity. For a wetting liquid, pulling the substrate out of the bath entrains a film. As a result of the competition between capillary forces and viscous dissipation, film thickness $h$ varies as $h_{LLD}\sim l_c Ca^{2/3}$ as predicted by Landau, Levich and Derjaguin (LLD)~\cite{landau_dragging_1942,derjaguin_thickness_1943}. The control parameter is the capillary number $Ca=\eta v/\gamma$ which compares viscous dissipation to capillary pressure gradients, where $\eta$ is the liquid viscosity and $v$ the pull-out velocity. This law has been verified experimentally for capillary numbers $Ca<0.01$~\cite{ruschak_coating_1985}.

\begin{figure}
\includegraphics[height=4.5cm]{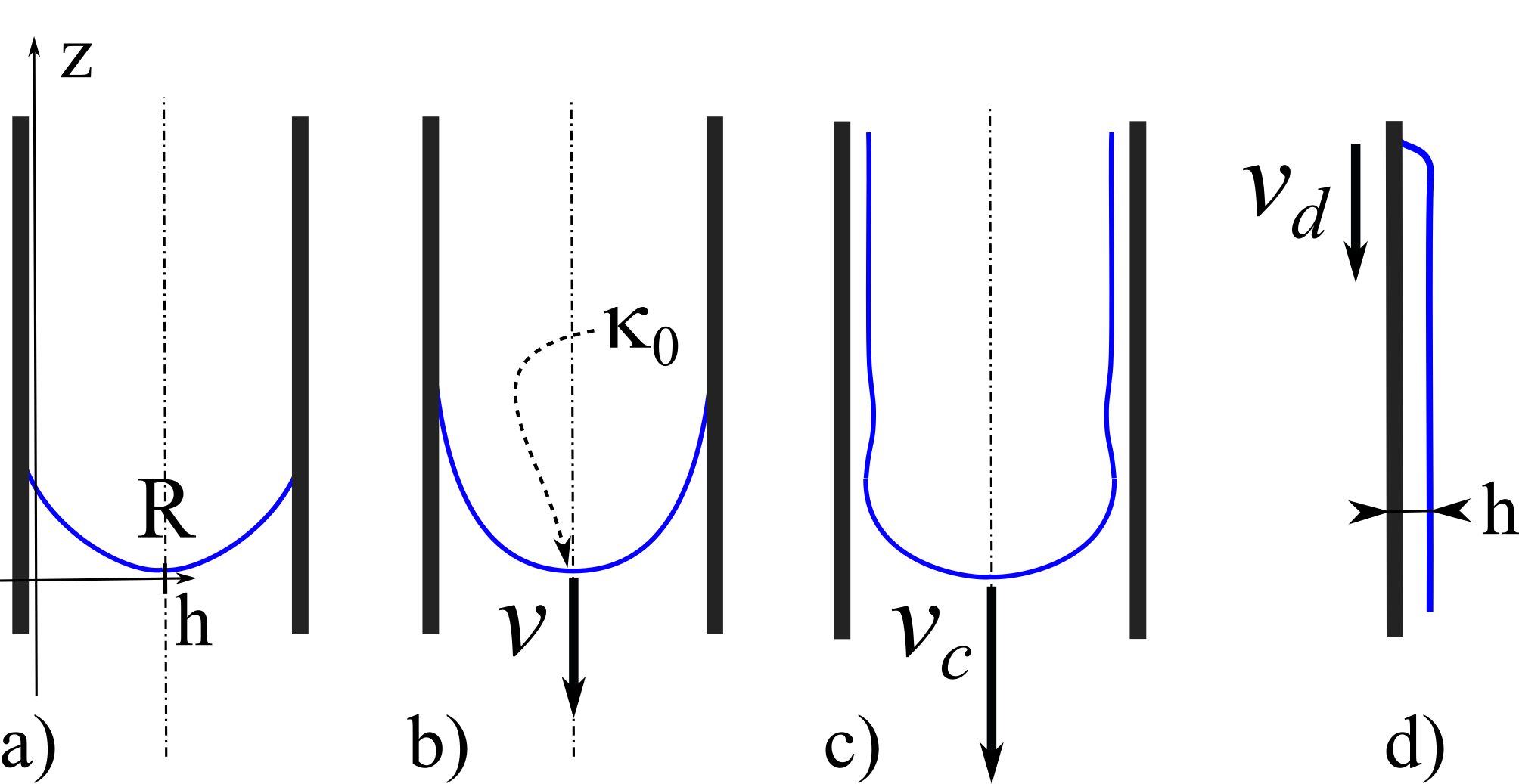}
\caption{\label{fig:dynamic_transition} Dynamic wetting transition for a partially wetting fluid in a tube, with gravity. At rest (a), the contact angle is finite. With velocity (b), the dynamic meniscus elongates and the contact angle decreases. At a critical velocity (c), the dynamic meniscus is non longer stable and a liquid film emerges. At its top (d), such a film terminates by a triple line dewetting at velocity $v_d$.}
\end{figure}

For partially wetting liquids, with non zero static contact angle (Fig.~\ref{fig:dynamic_transition} a), the dynamics is noticeably more complex. \added{When the substrate is pulled out of the liquid bath, hydrodynamic dissipation appears in the triple line region. At the macroscopic scale, the dynamic meniscus adopts an elongated shape and its contact angle decreases with velocity (Fig.~\ref{fig:dynamic_transition} b)~\cite{bonn_wetting_2009,snoeijer_moving_2013,rio2017withdrawing}. Above a critical velocity $v_c$, a dynamic meniscus with a steady state triple line is no longer stable and a liquid film is entrained (Fig.~\ref{fig:dynamic_transition} c). This destabilization of the dynamic meniscus, also called dynamic wetting transition, occurs at a} critical capillary number $Ca^*$ which depends on the equilibrium contact angle and is typically of the order of $10^{-3}$ to $10^{-2}$~\cite{petrov_existence_1985,snoeijer_avoided_2006,Zhao18a}. The concept of an intrinsic maximum dewetting velocity for the triple line has often been invoked to explain the transition from dynamic meniscus to film~\cite{blake_maximum_1979,petrov_existence_1985}. However, a more recent theoretical investigation of triple line dynamics based on lubrication theory proposed a different picture: beyond a critical velocity, the dynamic meniscus profile can no longer be matched to the static meniscus curvature~\cite{voinov_wetting:_2000,eggers_hydrodynamic_2004}. The critical velocity depends not only on the triple line dynamics but also on the macroscopic geometry. This prediction has to date not been validated experimentally.

Despite these complexities, film thicknesses for partially wetting liquids are expected to follow a very simple relation. \added{Dewetting with velocity $v_d$ feeds the film at the top contact line (Fig.~\ref{fig:dynamic_transition} d), and the flux $h v_d$ must match gravity driven drainage, with flux $\rho g h^3/(3\eta)$, resulting in a steady state, homogeneous film with thickness \begin{equation}\label{eq:film_thickness}h=l_c \sqrt{3Ca_d}\end{equation}
a result first predicted by Derjaguin~\cite{derjaguin_thickness_1943}.} However, Hocking~\cite{hocking_meniscus_2001} and Snoeijer~\cite{snoeijer_avoided_2006} have shown that within the lubrication approximation, given the triple line constitutive relation $\label{eq:ca_dynamics}\theta=\theta(Ca_d)$~\cite{blake_kinetics_1969,voinov_hydrodynamics_1976,cox_dynamics_1986,bonn_wetting_2009}, there is only one permitted value of triple line velocity $v_d$ (or capillary number $Ca_d$). As a result, for a partially wetting fluid, when a film forms, its thickness $h$ is unique and \emph{independent of the pull-out velocity $v$}. This simple but somewhat non-intuitive theoretical prediction has been confirmed numerically \cite{gao_film_2016} and experimentally by dip-coating silicone oils on fluorinated flat plates~\cite{snoeijer_thick_2008,delon_relaxation_2008} (with viscosities in the Pa.s$^{-1}$ range and Ca$^*\simeq 0.01$). However, the details of the forced wetting transition and how the dynamic triple line evolves into the film deposition regime remain unclear~\cite{snoeijer_avoided_2006}.

Here, we investigate film thickness selection for dewetting liquids with much lower viscosities, namely aqueous solutions on hydrophobic surfaces. To easily exceed the cm/s threshold velocity required for dynamic wetting, we drive aqueous bridges inside vertical polymeric tubes by application of pressure in a cylindrical geometry. To better understand the detailed morphology of the deposited films and how it evolves with time, we have developed an original experimental method. The resulting snapshots of deposited films reveal uniform thicknesses, but in contrast to previous results, we find that we can select the thickness of the steady state films. We show that  the film thickness and the triple line dynamics are still strongly interconnected, but that they can both be selected through the macroscopic curvature of the meniscus at the dynamic wetting transition. Thus, this thickness selection mechanism directly evidences the coupling between macroscopic meniscus geometry and meniscus stability advocated by Eggers~\cite{eggers_hydrodynamic_2004}. Conversely, it also points to more complex mechanisms for adjusting triple line dynamics to different film thicknesses at high capillary numbers.

\section{Experiments}

\begin{figure}
\includegraphics[height=5cm]{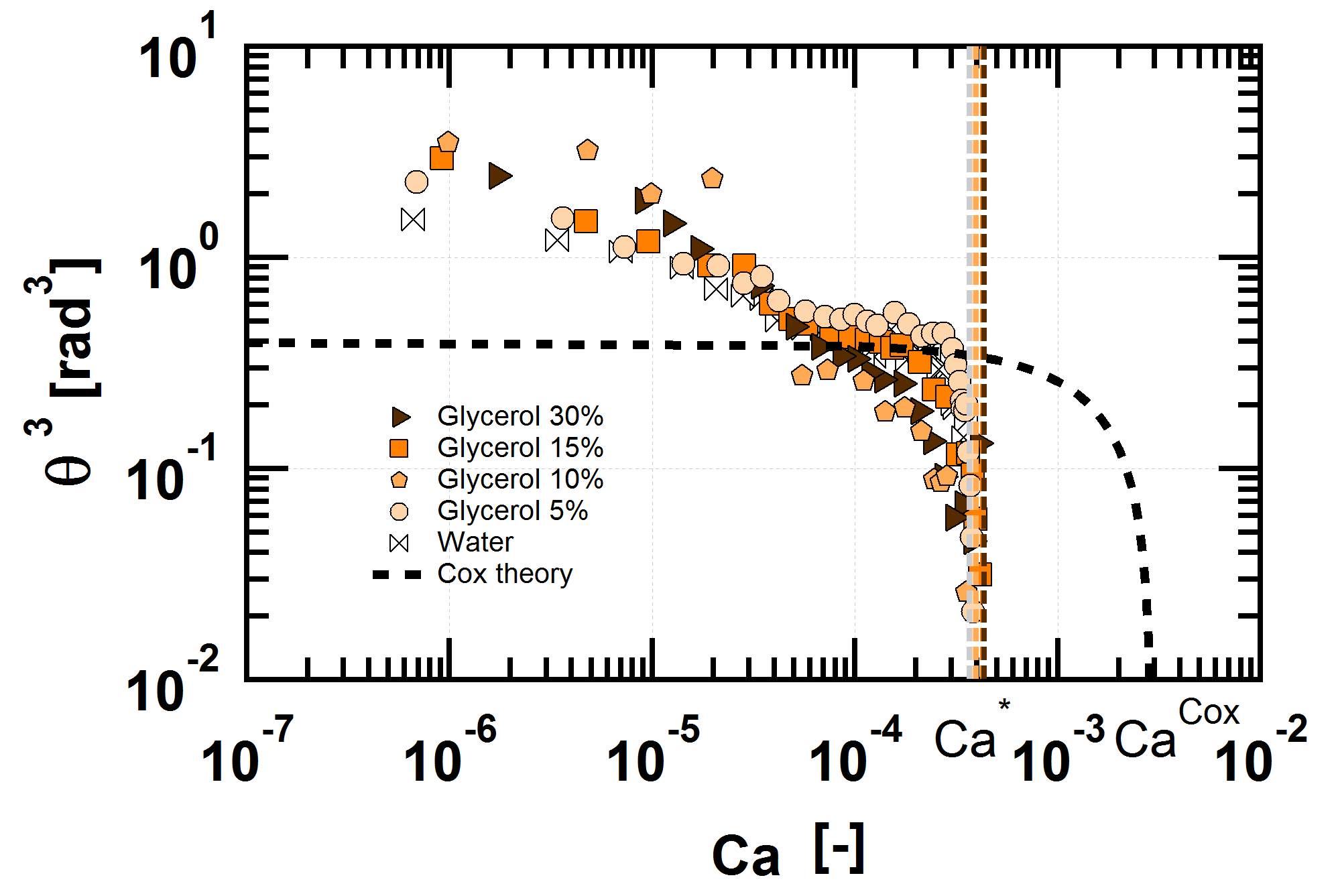}
\caption{\label{fig:Ca_slow} Receding contact angles as a function of velocity for different glycerol/water solutions at low velocities. The dashed line is a Cox-Voinov law. Despite some scatter, the evolution is faster than predicted by purely viscous dissipative models. Around $Ca$=3$\ $10$^{-4}$ there is a rather sharp transition towards the dynamic wetting regime where a thin liquid film is deposited.}
\end{figure}

\begin{figure}
\includegraphics[height=5cm]{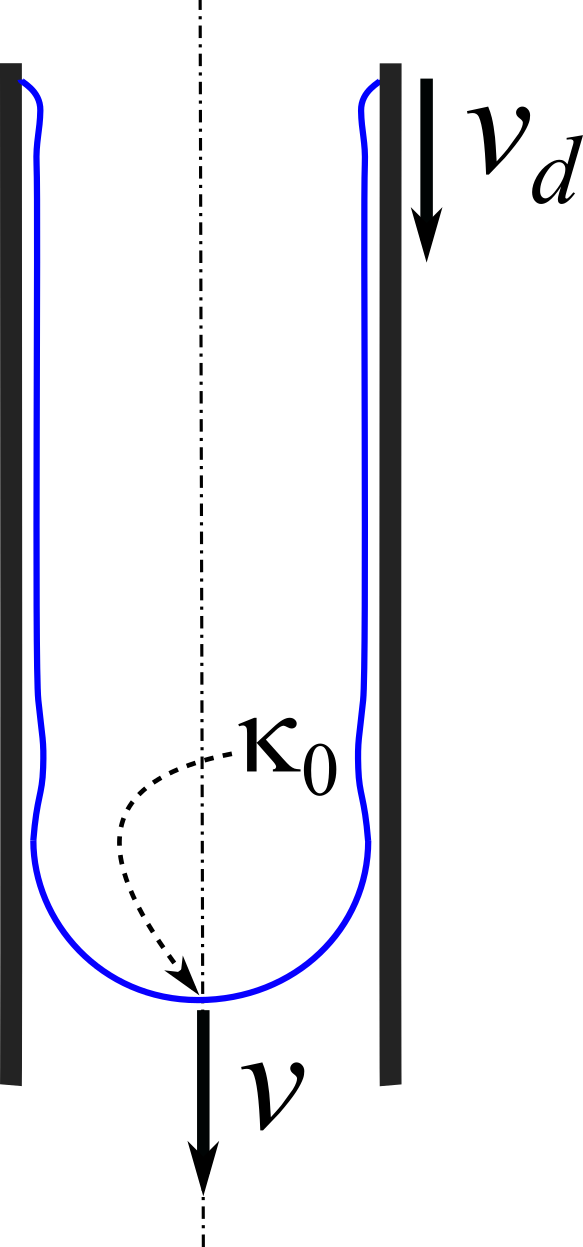}
\includegraphics[height=5cm]{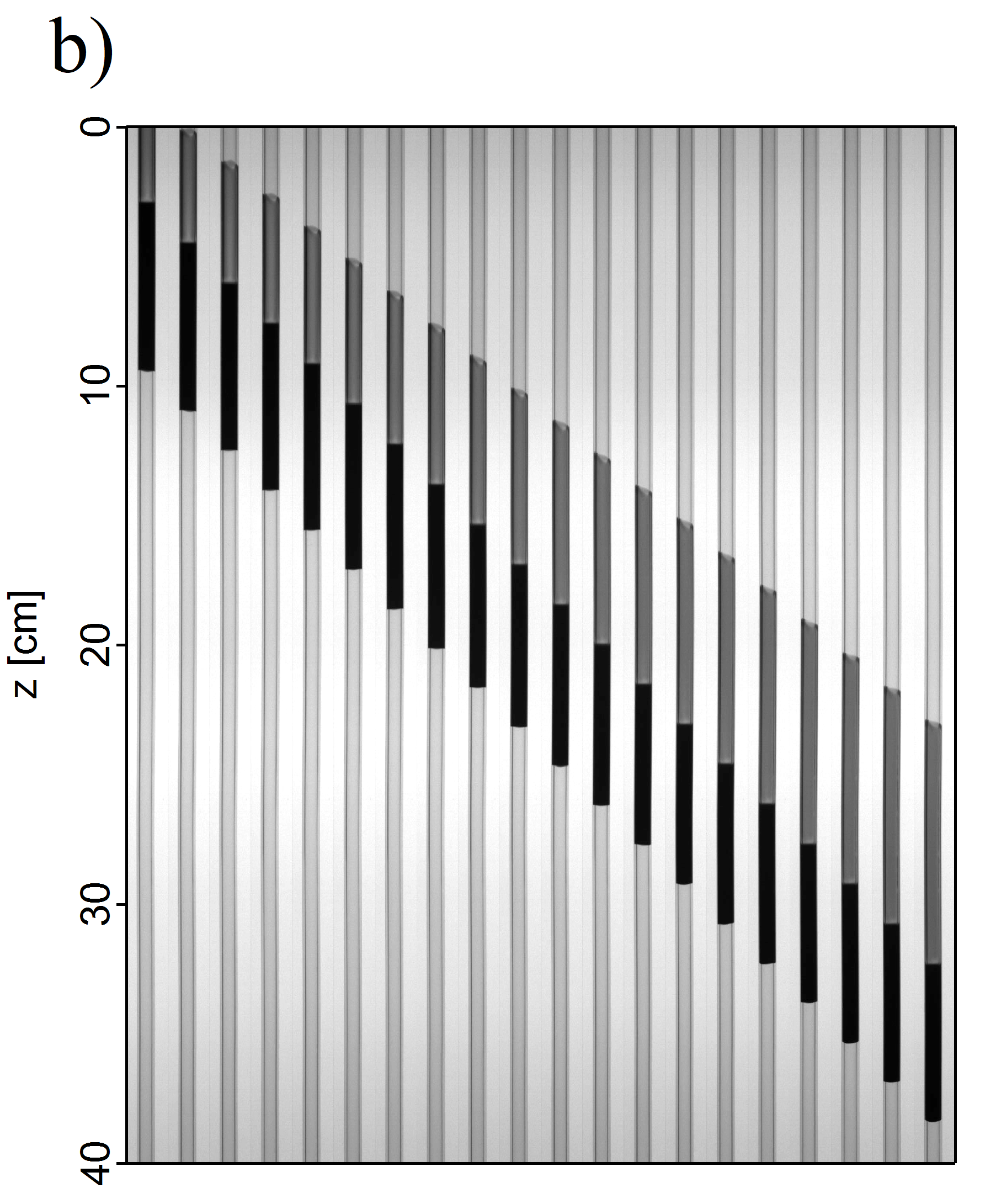}
\includegraphics[height=5cm]{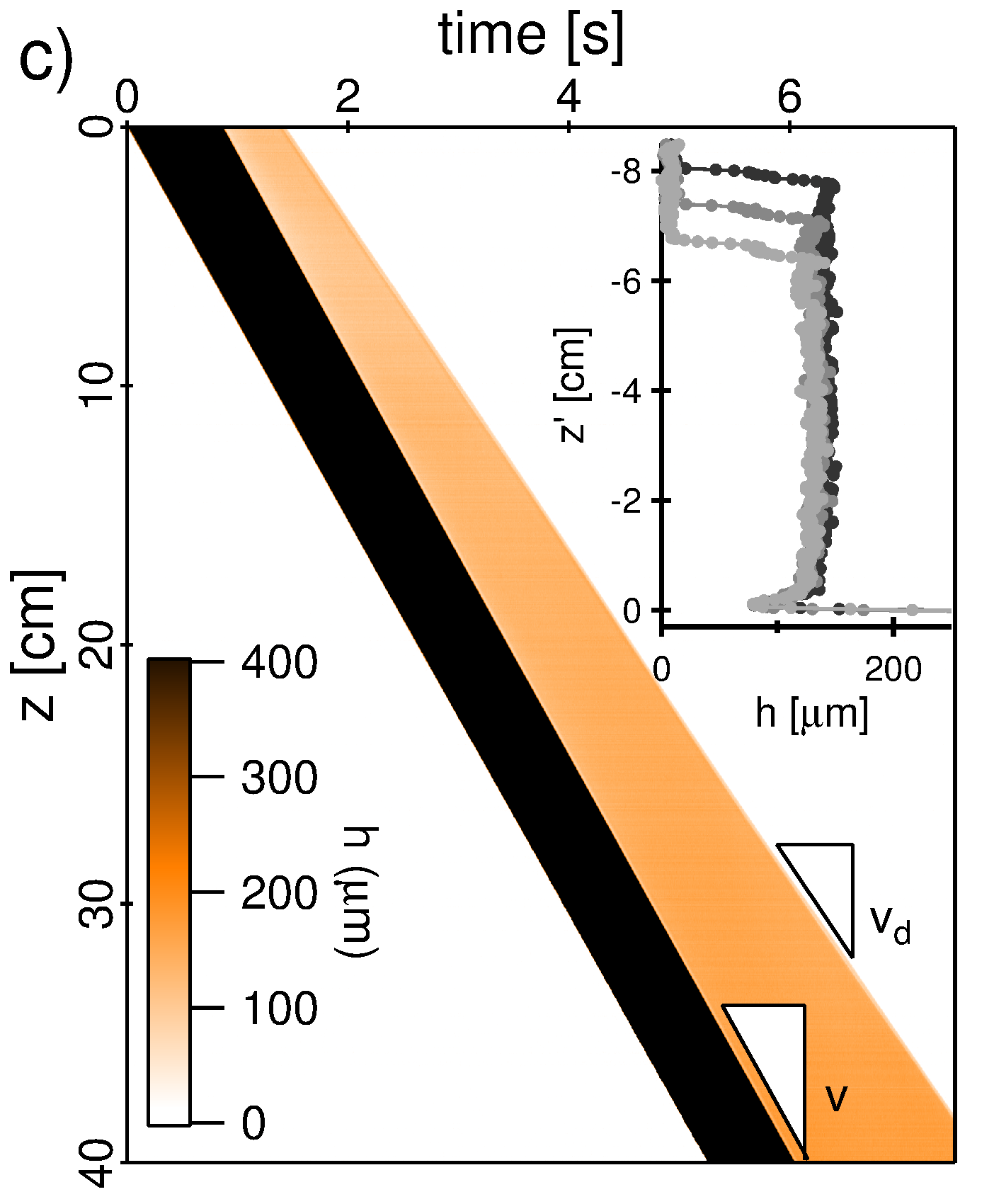}
\caption{\label{fig:spatio}a) schematics of liquid meniscus and matching film in the dynamic wetting regime; b) a series of 20 snapshots taken at time interval 0.2 s while the liquid bridge (dark) is flushed down the tube (light grey) (Ca=1.9e-3). The liquid is a 20\%wt glycerol solution in water. The medium grey level is the liquid film and the space and time variations of the film thickness can be measured after calibration of the optical absorption ; c) space-time plot showing film thickness (h=132 $\mu m$); inset: three successive thickness profiles at 0.5 s time increment, shown in the reference frame of the bridge. There is a small velocity difference between the film dewetting velocity $v_d$ at the top and the bridge velocity $v$ at the bottom of the film which is seen to extend slowly at constant thickness.}
\end{figure}

Fast dewetting experiments were performed in vertical tubes with 1.5 meter length and 6.4 mm inner diameter. The tube material was plasticized polyvinylchloride~(Tygon, Saint-Gobain). A 2~mL bridge of liquid was introduced at the top of the tube. Subsequently, pressurized nitrogen was allowed at the top so that the liquid bridge, which was initially at rest, reached a stationary velocity within a transient time negligible compared to the total measurement duration. For low velocity experiments, a similar setup, smaller in size, was also used. This allowed to index match the outside of the tube in a containing cell with parallel faces for better optical observations. Steady state velocities ranged between $10^{-6}$ and 1~m.s$^{-1}$ for overpressures of the order of tens of Pa.

We used solutions of glycerol in water with mass fractions ranging from 0 to 75 wt\%. Viscosity increased from 0.94 to 30 mPa.s and surface tension decreased from 70.3 to 59.0 mN.m$^{-1}$ with glycerol addition. With these combinations of surface tensions and viscosities, the capillary length was nearly constant, varying from 2.66~mm to 2.25~mm, close to the tube radius $R$=3.2~mm, while the characteristic velocities of the liquids, defined as $v_l=\gamma/\eta$, ranged from 2 to 75~m.s$^{-1}$.

Quantitative absorption imaging provided snapshots of film thickness profiles over a 40~cm field of view along the tube. To that end, $4.9~10^{-4}\ $mol.L$^{-1}$ New Coccine, a red dye, was added to the liquids. This dye was selected so that viscosity, surface tension, and interfacial tension were not affected, and no adsorption nor absorption of the dye was detected after prolonged contact with the tubes. A white LED screen provided backlight illumination and green-filtered images (10~nm bandwidth, 508~nm central wavelength for maximum absorption) were collected with a camera. Absorbance was calibrated against precision rectangular cells with 10 to 800~$\mu$m thickness. Images were captured at frame rates between 5 and 1000 Hz. Intensity from the vertical middle section was converted into film thickness $h$ by the Beer-Lambert law $h=-1/(2\epsilon c)\ln{(I/I_0)}$ where $I$ is the transmitted light intensity, $I_0$ is the intensity transmitted through a non absorbing water film, $\epsilon =44945 \pm 40$~L.mol$^{-1}$cm$^{-1}$ is the measured extinction coefficient and $c$ is the dye concentration. This method provided instantaneous profiles of film thickness with a 5~$\mu$m accuracy.  It improves over most tube experiments reported in the literature, where homogeneous and constant film thicknesses are assumed and evaluated as averages based on mass conservation~\cite{aussillous_quick_2000,callegari_dewetting_2005}. A few experiments were carried out at a lower dye concentration so that an image of the full meniscus shape could be recorded to evaluate the variation of its curvature.
\section{Results}
The macroscopic receding contact angles $\theta$ of the dynamic meniscus were measured at low velocities (Fig.~\ref{fig:Ca_slow}). For all the fluids we observe an initial decrease of the receding contact angle followed, despite some significant scatter in the data, by a more or less extensive plateau-like behaviour with a dynamic contact angle approximately equal to 40$^\circ$. This plateau is followed by an abrupt decrease and $\theta$ tends to zero for $Ca^\star\simeq 3~10^{-4}$. The data are compared to the Cox-Voinov prediction for a receding triple line \begin{equation}\label{Eq:Cox-Voinov}\theta^3-{\theta_e}^3\simeq 9 Ca \ln\left(\frac{l_c}{a}\right)\end{equation}
\noindent where $\theta_e$ is the equilibrium contact angle and $l_c$ and $a$ are the upper and lower cut-off lengths, with $\ln({l_c}/{a})\simeq 15$. Setting $\theta_e=40^{\circ}$ in Eq.~\ref{Eq:Cox-Voinov}, the plateau is fitted to the Cox-Voinov law (dashed line) which is found to overestimate $Ca^\star$ since forced wetting is predicted to occur for $Ca^\star\simeq 3~10^{-3}$, one order of magnitude larger than measured. Note that there is an intermediate velocity range between $Ca \sim 2~10^{-4}$ and $Ca^*$ (Fig.~\ref{fig:Ca_slow} - dashed area) where 20 cm long tube segments are too short to establish a steady state regime with confidence, because the characteristic times of the transients become increasingly large). In fact, using the long tubes, for all the liquids tested, forced wetting is observed above $Ca^*\sim 4~10^{-4}$ and a liquid film forms almost instantaneously (Fig.~\ref{fig:spatio}b). Space-time plots of film profiles are built from image series (Figure~\ref{fig:spatio}c). The dark area corresponds to the liquid bridge, the orange area marks the presence of the liquid film and the white area is the bare tube. Three thickness profiles at different times are also plotted in Fig.~\ref{fig:spatio}c (inset). For series of snapshots, the rear meniscus velocity $v$ and the contact line velocity $v_d$ could be measured precisely (Fig.~\ref{fig:spatio}a, c).

\begin{figure}
\includegraphics[height=4.5cm]{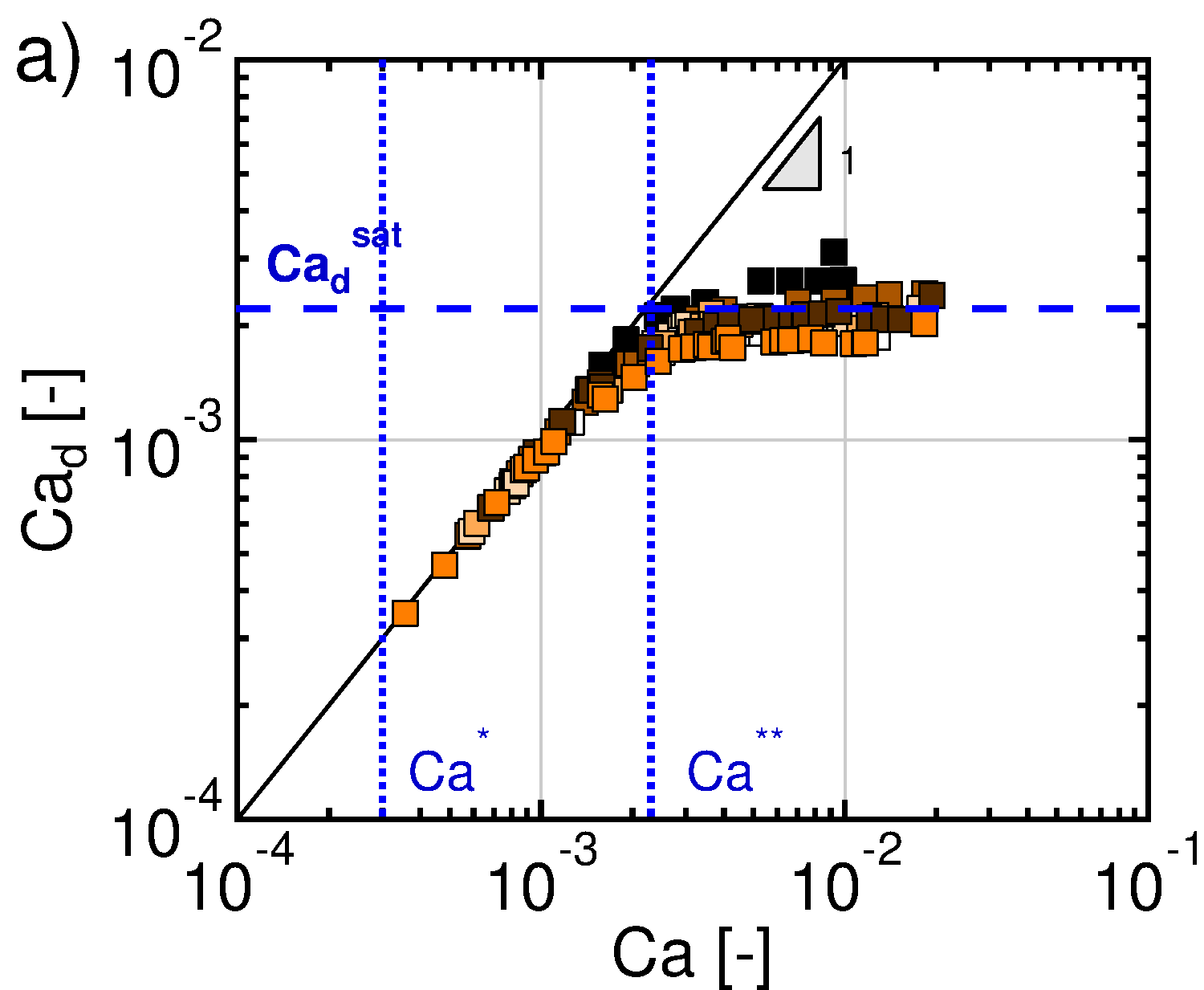}
\includegraphics[height=4.5cm]{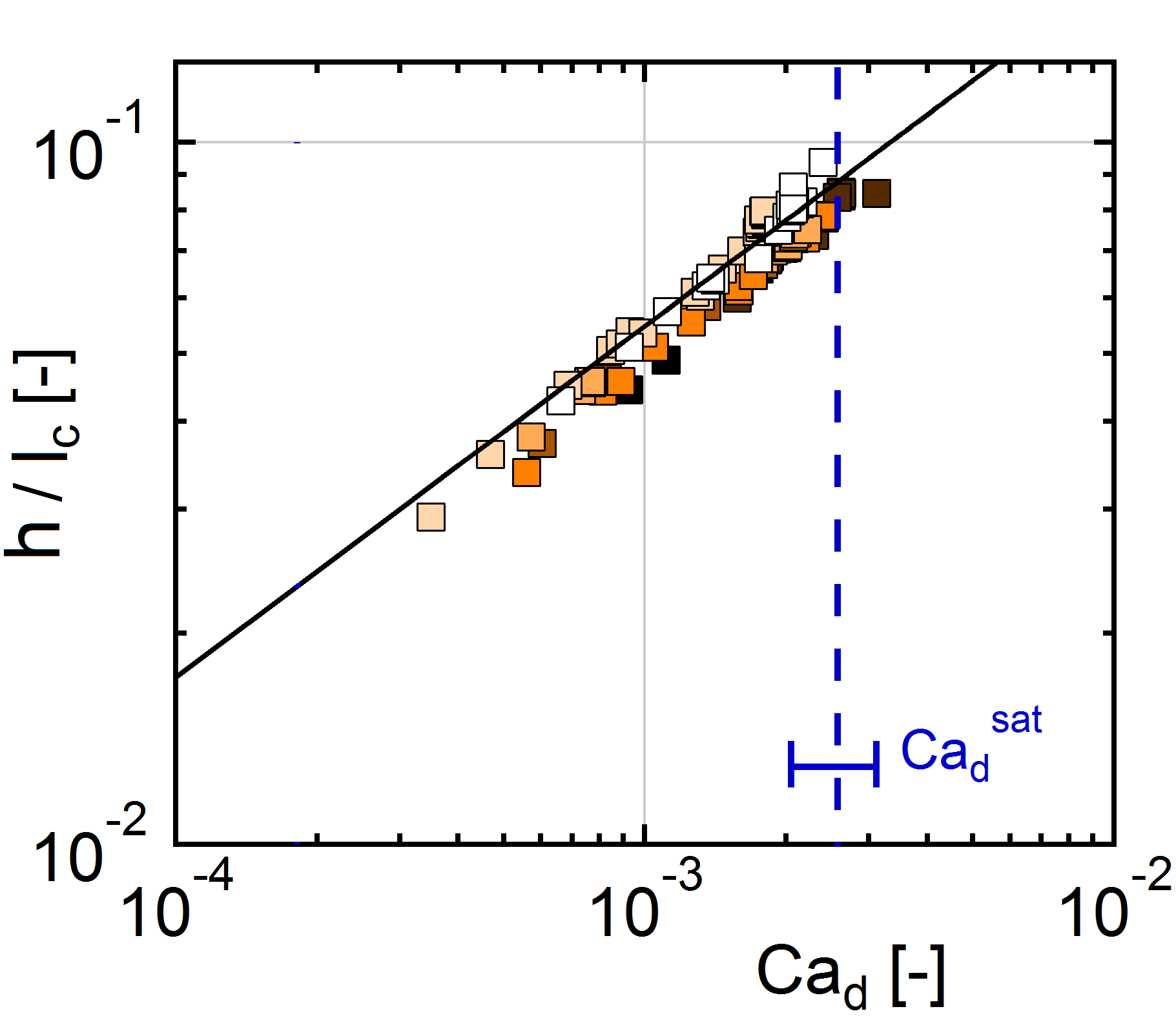} \includegraphics[height=4.5cm]{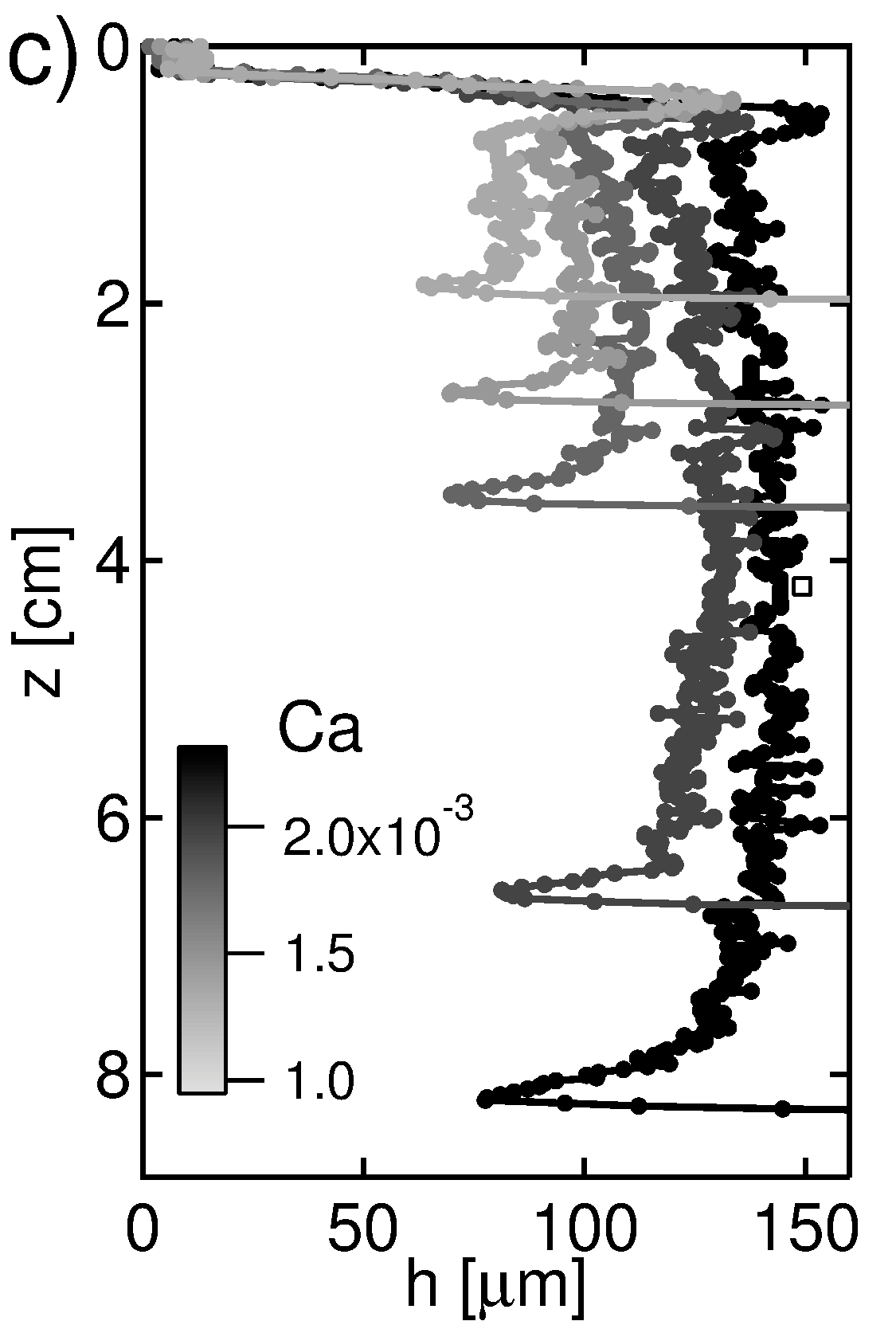}
\caption{\label{fig:exp:profiles:Ca} a) dewetting velocity $Ca_d$ as a function of meniscus velocity $Ca$ showing a
nearly proportional regime in the range $Ca^*$ to $Ca^{**}$, followed by saturation; b) normalized film thickness as a function of dewetting velocity and the prediction of Eq.~\ref{eq:film_thickness} (solid line); c) film thickness profiles for different $Ca$ between 9.6$\ $10$^{-4}$ and 2.3$\ $10$^{-3}$ plotted in the reference frame of the triple line - note the dimple at the junction between film and meniscus, and the rim near the triple line. a,b) glycerol:water from 0 to 75\%wt from light to dark; c) 20\%wt glycerol:water solution. For $Ca^*<Ca<Ca^{**}$, the film is directly connected to the meniscus: in this regime, we find that the film thickness can be selected through the meniscus velocity, in contrast to literature results~\cite{gao_film_2016,snoeijer_thick_2008,delon_relaxation_2008}}
\end{figure}

We observe that the film profiles exhibit a nearly uniform thickness, and that this thickness stays constant over time (Fig.~\ref{fig:spatio}b, c). Important additional features are what appears to be a shallow rim at the top and a dimple at the junction with the meniscus (Fig.~\ref{fig:exp:profiles:Ca}c). The dewetting velocity $v_d$ at the triple line is slightly smaller than the meniscus velocity $v$ so that the film slowly grows in length at constant thickness. These results clearly show that the flows at both ends of the film are steady state (Fig.~\ref{fig:spatio}a) but as anticipated by Hocking~\cite{hocking_meniscus_2001}, the capillary number at the triple line $Ca_d$ is slightly smaller than the capillary number of the meniscus $Ca$. Figure~\ref{fig:exp:profiles:Ca}a shows that for $Ca>Ca^{*}$, and for various liquids and forcing conditions, $Ca$ and $Ca_d$ are related by a slightly sublinear relation until a second threshold denoted $Ca^{**}$ is reached. This upper limit will be explained in more detail in the discussion.
The measured film thickness $h$, normalized by capillary length $l_c$, is plotted as a function of $Ca_d$ in Fig.~\ref{fig:exp:profiles:Ca}b over the same range.

\begin{figure}
\includegraphics[height=5cm]{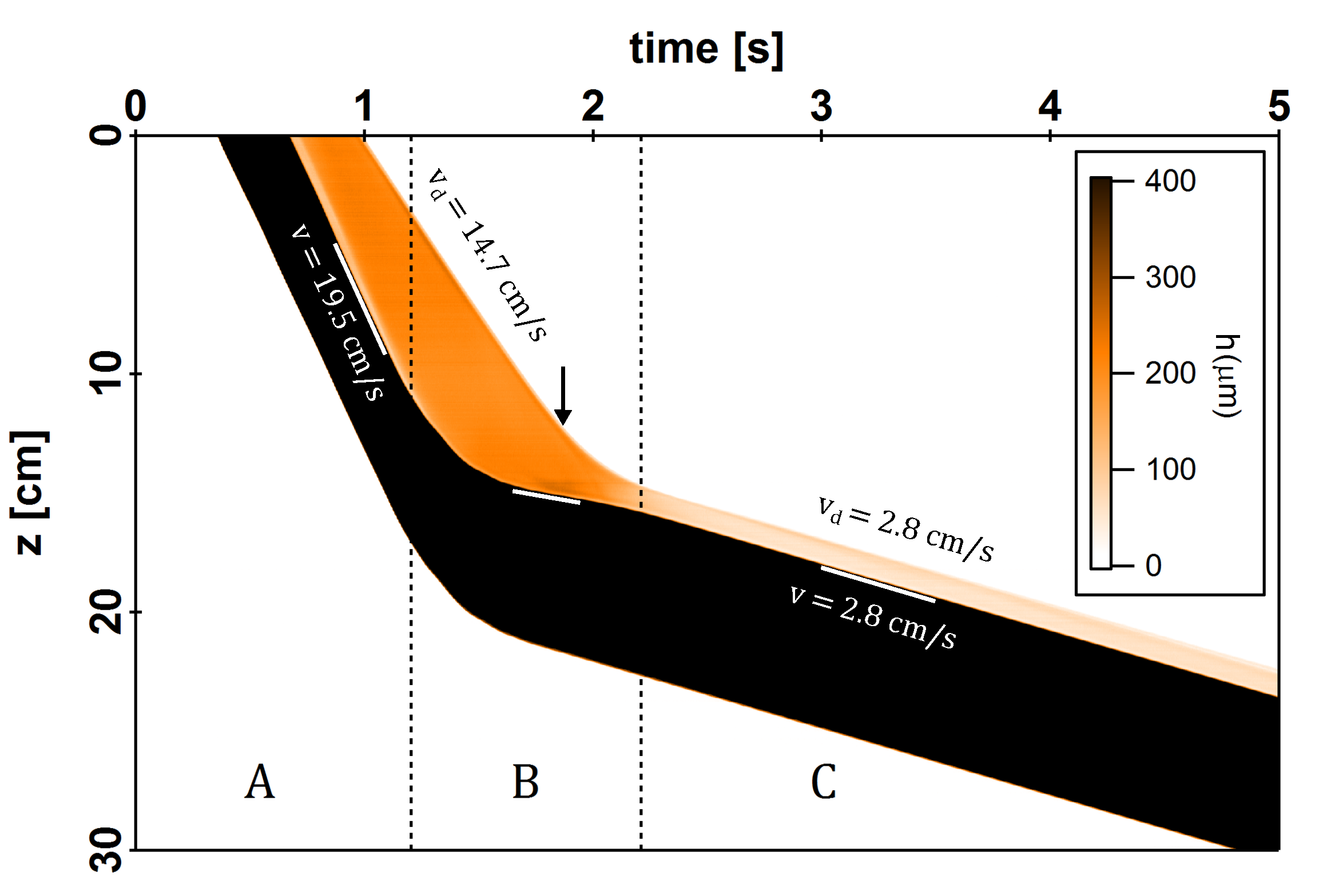}
\caption{\label{fig:instationary}Space-time plot of film thickness profile in a sudden deceleration experiment. From steady state, pressure is abruptly reduced to near threshold value: the bridge slows down precipitously (from $Ca=2.6\ 10^{-3}$ to $3.8\ 10^{-4}$) but the film thickness and dewetting velocity remain unaffected until the film almost disappears. This experiment directly demonstrates how once formed, the film is effectively decoupled from the meniscus. The liquid is pure water.}
\end{figure}

The remarkable result, which contrasts with other observations in the literature~\cite{snoeijer_thick_2008}, is that the film thickness varies significantly, over one order of magnitude. The Derjaguin relation between $Ca_d$ and $h$ Eq.~\ref{eq:film_thickness} is plotted in Fig.~\ref{fig:exp:profiles:Ca}b with no adjustable parameter and the agreement is excellent. The LLD relation was tested against the data: $h/l_c$ does not follow a $Ca^{2/3}$ law. This result suggests that the steady state film thickness {\em is} controlled by dewetting at the top of the film, as expected from~\cite{hocking_meniscus_2001,snoeijer_thick_2008}, and not by the dynamic meniscus at the bottom, as it would be in the LLD regime. 
This decoupling between film and meniscus was further tested through more elaborate experiments. After dynamic film formation at a given $Ca>Ca^*$ (Fig.~\ref{fig:instationary}, zone A), we swiftly decrease pressure. The bridge slows down precipitously (zone B) and finds a new steady state velocity (zone C). We observe no concomitant variation of film thickness $h$ or dewetting velocity $v_d$ until the film, which now recedes much faster than the meniscus, has shrunk to a new equilibrium. Such transients clearly demonstrate the capacity of the film-meniscus junction to accommodate a set film thickness at different velocities, so that $v_d$ is independent of $v$. 
We conclude that triple line velocity and meniscus velocity are decoupled, since once the triple line velocity has been set, it is retained afterwards, also setting the film thickness.

However, we also find that we can select the velocity and thickness, in contrast to previous results~\cite{hocking_meniscus_2001,snoeijer_thick_2008}. Therefore, this selection mechanism can only operate {\em at film birth}, \emph{i.e.} upon destabilization of the \added{dynamic meniscus}. To better understand this process, we now numerically investigate the stability of \added{dynamic menisci} in tubes, \added{\emph{i. e.} we explore the nature of steady state solutions as a function of both velocity and applied pressure, as well as the conditions under which solutions with triple lines exist.}
\section{Model}

\begin{figure}
\includegraphics[height=6cm]{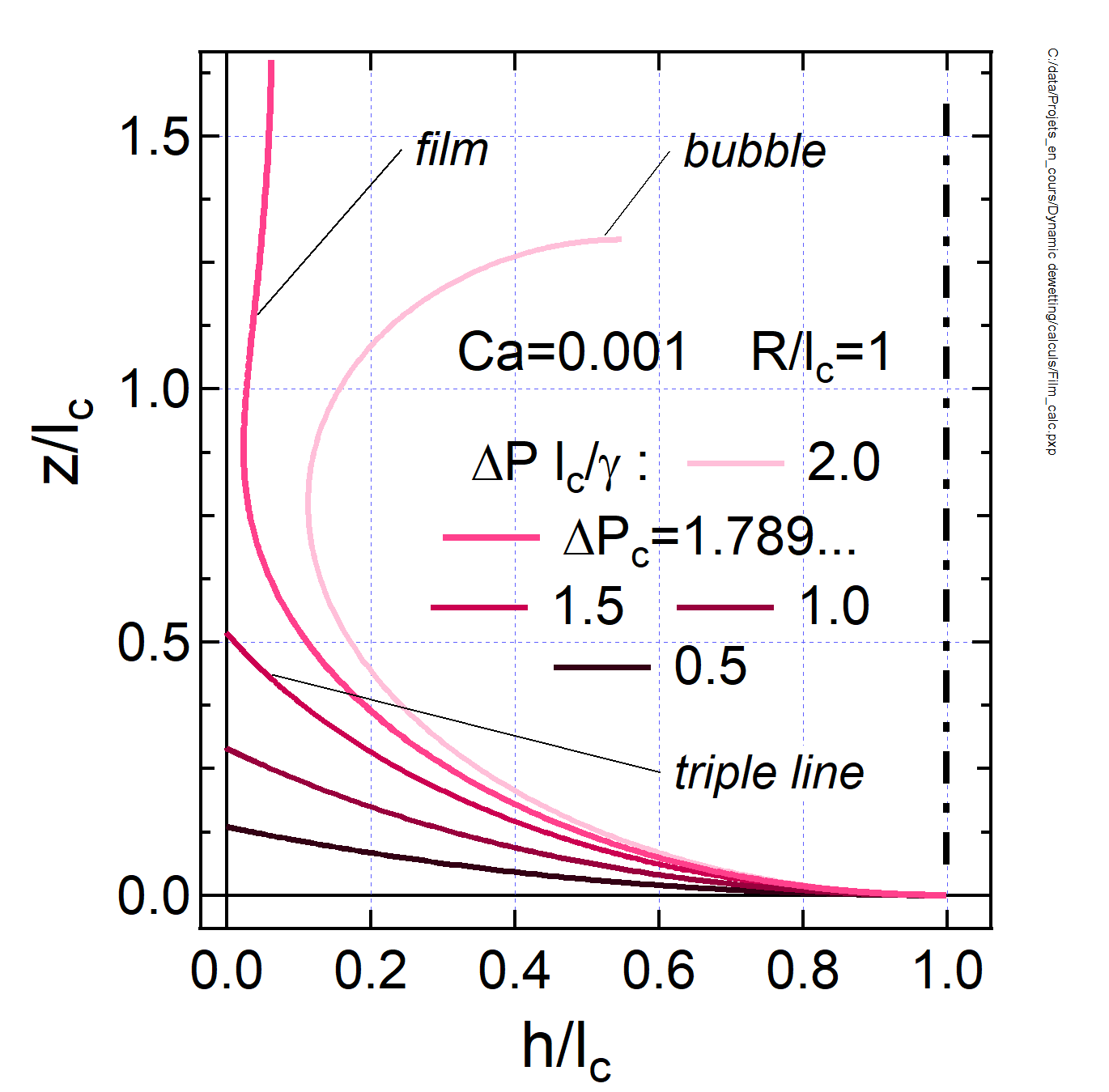}

\caption{\label{fig:h_z_DP_Ca_p001} Steady state solutions to Eq.~\ref{eq:h_steady_state} with zero flux ({\em i.e.} $\textrm{cte}=0$) for Ca=$0.001$ and $R/l_c=1$ for various pressure values spanning the dynamic wetting transition. For pressures below the critical pressure $\Delta P_c$, the meniscus solutions meet the tube surface with a dynamic triple line. At $\Delta P_c$, the meniscus solution ends into a thin film with zero total flux. For  $\Delta P > \Delta P_c$ the solutions are bubbles which are not relevant for our system.}
\end{figure}

We calculate steady state profiles $h(z)$ for an axisymmetric, dynamic meniscus moving down a vertical tube in the presence of gravity, \added{with the geometry shown in Fig.~\ref{fig:dynamic_transition} b}.  The notable difference between a plate and a tube is that for the tube the applied pressure $\Delta P$ sets the macroscopic meniscus curvature $\kappa_0=\Delta P/\gamma$. Thus pressure enters the calculations as the curvature boundary condition on the symmetry axis.  \added{As will be shown in detail below, depending on pressure, three types of solution may appear, which are shown in Fig.~\ref{fig:h_z_DP_Ca_p001}. The end point of the profile may lie at the inner surface of the tube ($h=0$), in which case we have a dynamic meniscus with triple line. Another possible case is a profile which does not touch the tube surface but curves back into the liquid, thus forming a Bretherton-type bubble~\cite{bretherton_motion_1961}. Finally, at the boundary between these two cases, we find a limit case where the liquid surface ends up parallel to the tube wall at large positive $z$ values, \emph{i.e.} a liquid film.}

More precisely, within the standard model, using the lubrication approximation~\cite{voinov_wetting:_2000,eggers_toward_2004}, in the meniscus frame, the profile obeys
\begin{equation}\label{eq:h_steady_state}\frac{h^3}{3\eta} ( \gamma\partial_z \kappa + \rho g ) - v h = \textrm{cte}\end{equation}
where surface curvature is given by
\begin{equation}\label{eq:kappa_steady_state}\kappa = -\frac{\partial^2_{zz} h}{\left(1+(\partial_z h)^2\right)^{3/2}}+\frac{1}{(R-h)\sqrt{1+(\partial_z h)^2}}\end{equation}
In Eq.~\ref{eq:kappa_steady_state}, the second term characterizes the axisymmetric geometry and is \emph{e.g.} responsible for the Plateau instability of liquid columns.

The constant $\textrm{cte}$ in Eq.~\ref{eq:h_steady_state} is set by the incoming flux: because we model a steady state triple line (Fig.~\ref{fig:dynamic_transition} b), the total flux is zero and $\textrm{cte}=0$. To handle the triple line singularity, we introduce a small slip length $\lambda$ which is taken equal to $10^{-4} l_c$ unless otherwise noted.

\added{Note also that Eq.~\ref{eq:h_steady_state} is a third order differential equation. We shoot from bottom of the meniscus, at $z=0$ on the symmetry axis. the boundary conditions are initial thickness ($h=R$), slope ($h'^{-1}=0$) and curvature (given by $\Delta_P$). If the solution meets the tube ($h(z)=0$ for some positive $z$), then there is a triple line. If not, the solution is a bubble. Note that in this numerical scheme, the microscopic contact angle $\theta_{0}$ is not yet a boundary condition: it is result of the calculation, and a function of the parameters $\Delta_P$, $Ca$ and $R$. We can of course prescribe a microscopic contact angle, \emph{e. g.} with a specific constitutive relation specifying the microscopic contact angle as a function of velocity, for a given liquid/surface couple. As will be shown below (Sec.~\ref{Sec:phase_diagram}), with this condition we determine the pressure $\Delta P$ which has to be applied to obtain a meniscus with steady state velocity $Ca$.}
\subsection{Numerical results}
We have computed meniscus profiles for different velocities and pressures for a tube radius $R=l_c$, close to our experimental conditions. All spatial dimensions are normalized by $l_c$ including film thickness $h$, vertical coordinate $z$ and tube radius $R$ which plays the role of a Bond number since $Bo=(R/l_c)^2$. The pressure $\Delta P$ is normalized by $\gamma/l_c$  and thus equal to the normalized curvature. The normalized forms of Eqs.~\ref{eq:h_steady_state} and \ref{eq:kappa_steady_state} can be found in Appendix~\ref{app:norm_eq}.

For $Ca=10^{-3}$, several profiles are plotted in Fig.~\ref{fig:h_z_DP_Ca_p001} for increasing $\Delta P$. At low pressures, the curvatures are small and the profiles meet the tube wall thereby forming a dynamic triple line with a microscopic contact angle defined by $\tan\theta_0=h'$ where $h=0$. This microscopic contact angle decreases with increasing pressure (curvature). Because of the low value of the capillary number, the dynamic effects near the triple line are very limited. In particular, variations of the slip length do not elicit significant impact on the profile except for finer details in the immediate vicinity of the triple line (not shown). For higher values of $\Delta P$, however, the curvature becomes large enough that the profile no longer meets the surface and we find the bubble solutions of the Bretherton type~\cite{bretherton_motion_1961}, which are not relevant here. The limit between these two cases, where the dynamic meniscus is matched to a flat (steady state, zero flux) film solution, is obtained for a well defined critical pressure $\Delta P_c$ which we now investigate in more detail.
\begin{figure}
a)\includegraphics[height=6cm]{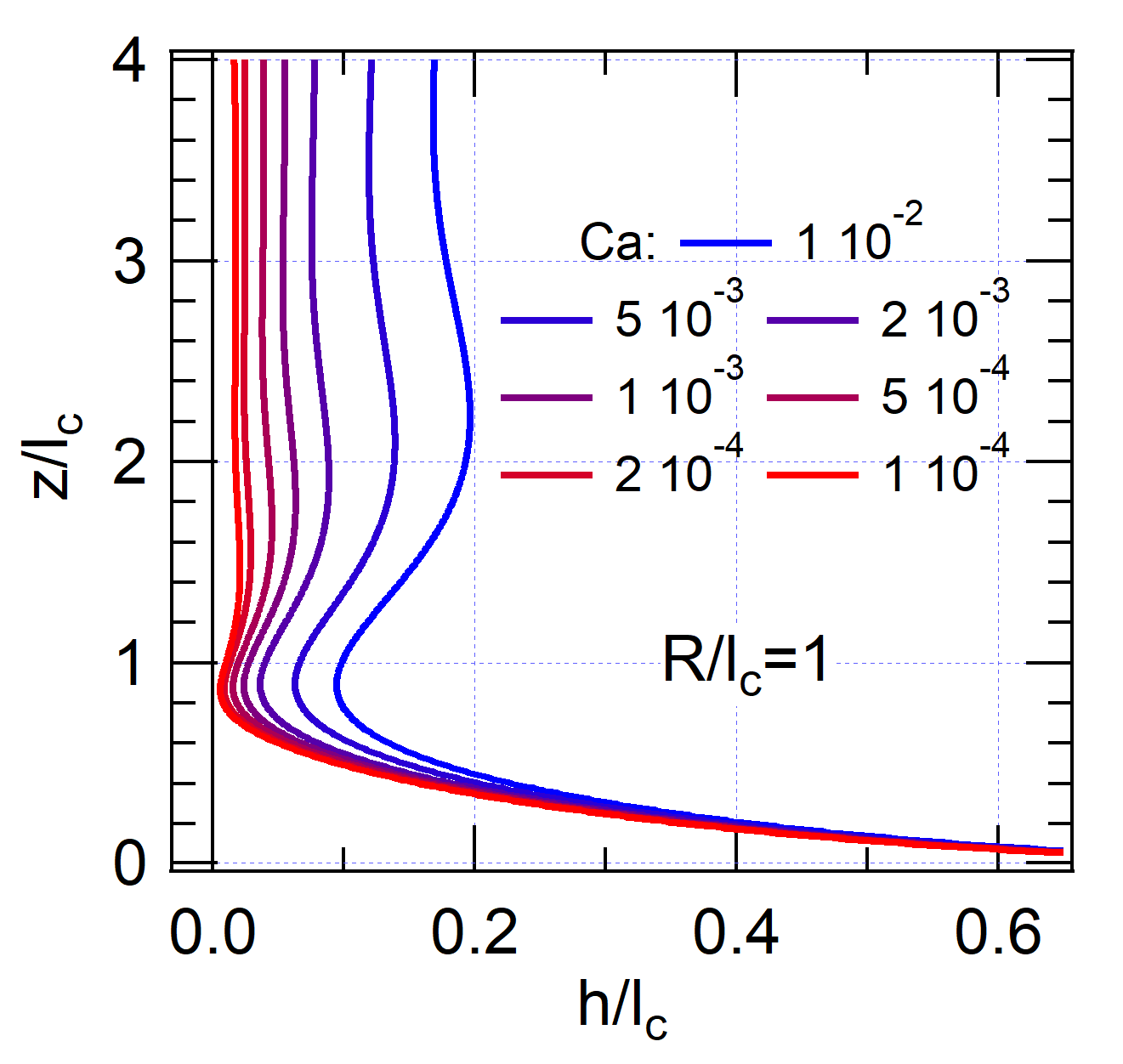}
b)\includegraphics[height=6cm]{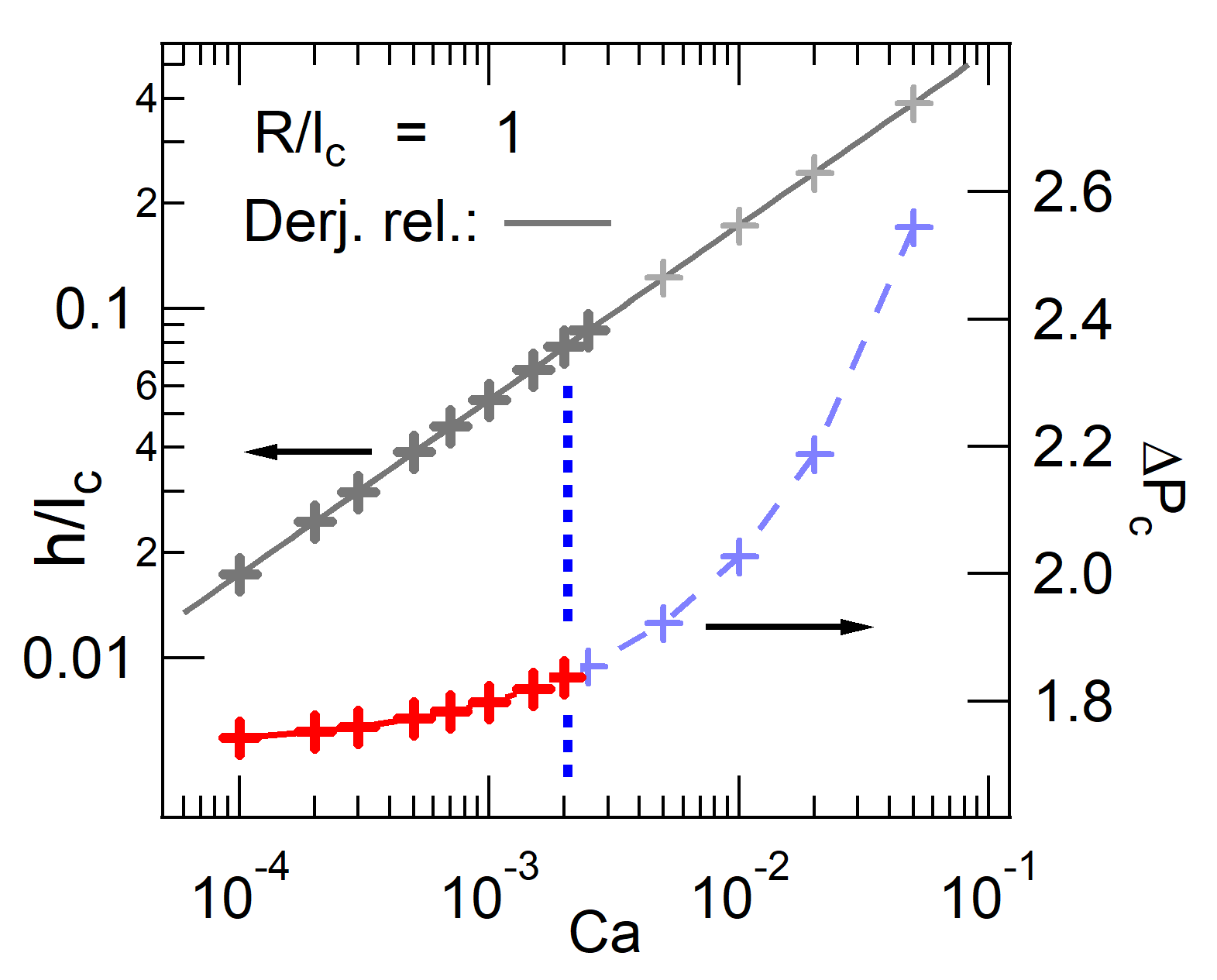} 
\caption{\label{fig:numerics:Ca} a) Steady state solutions with a zero flux film for $R/l_c=1$ and different values of $Ca$. Each velocity elicits a different value of the critical pressure $\Delta P_c$. The film thickness increases with velocity, while the oscillatory nature of the solution becomes more and more apparent; b) critical pressures and corresponding film thicknesses for different meniscus velocities $Ca$. The predictions of Eq.~\ref{eq:film_thickness} between film thickness and $Ca$ match the numerical results. The weak dependence of the critical pressure on $Ca$, especially in the Ca range of interest for our experimental system ($Ca < 2\ 10^{-3}$) shows that minute increases in pressure result in significant thickness variations.}
\end{figure}

We have computed this critical pressure for different values of velocity $Ca$, along with the corresponding liquid film profiles $h(z)$ (Fig.~\ref{fig:numerics:Ca}a). The film thickness increases as the velocity increases and the oscillatory nature of the solution becomes more and more apparent. The relation between film thickness $h$ and $Ca$ is plotted in Fig.~\ref{fig:numerics:Ca}b (left hand axis): it obeys Eq.~\ref{eq:film_thickness} with no adjustable parameter. This is no surprise since by definition of the triple line, we impose a zero flux condition as the film forms. With this and a flat film, Eq.~\ref{eq:h_steady_state} simplifies into Eq.~\ref{eq:film_thickness}. The relation between critical pressure $\Delta P_c$ and $Ca$ is plotted in Fig.~\ref{fig:numerics:Ca}b (right hand axis). We find a definite variation of the critical pressure with velocity, although it remains moderate in the range $Ca < Ca^{**}=2\ 10^{-3}$ which is of primary interest in our experiments. In particular, $\Delta P_c$ converges to a finite value $\Delta P_{c0}$ as the velocity goes to zero. These results evidence how the mechanism for film thickness selection is related to macroscopic curvature and pressure.

\begin{figure}
a) \includegraphics[height=5cm]{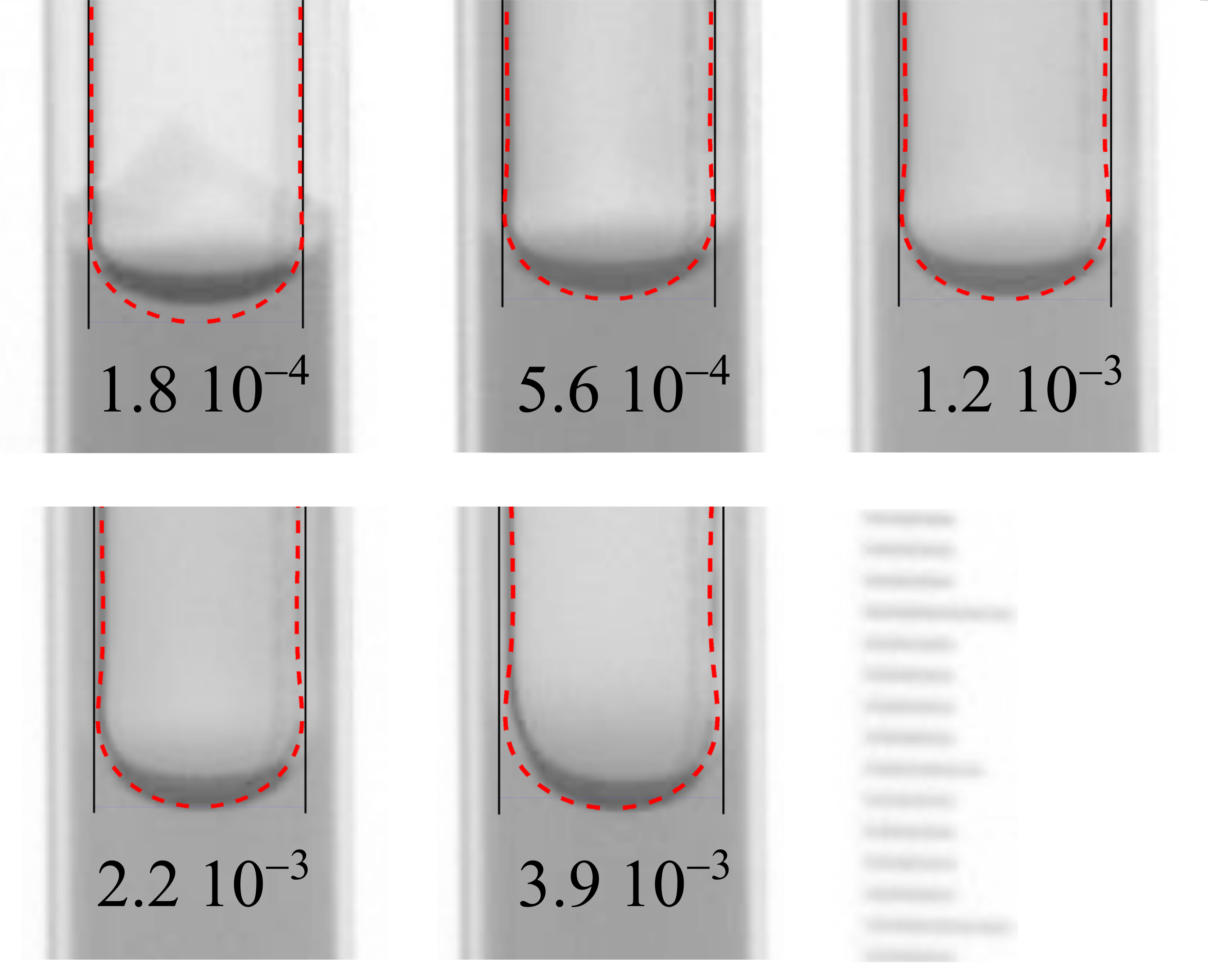}\linebreak
b) \includegraphics[height=4cm]{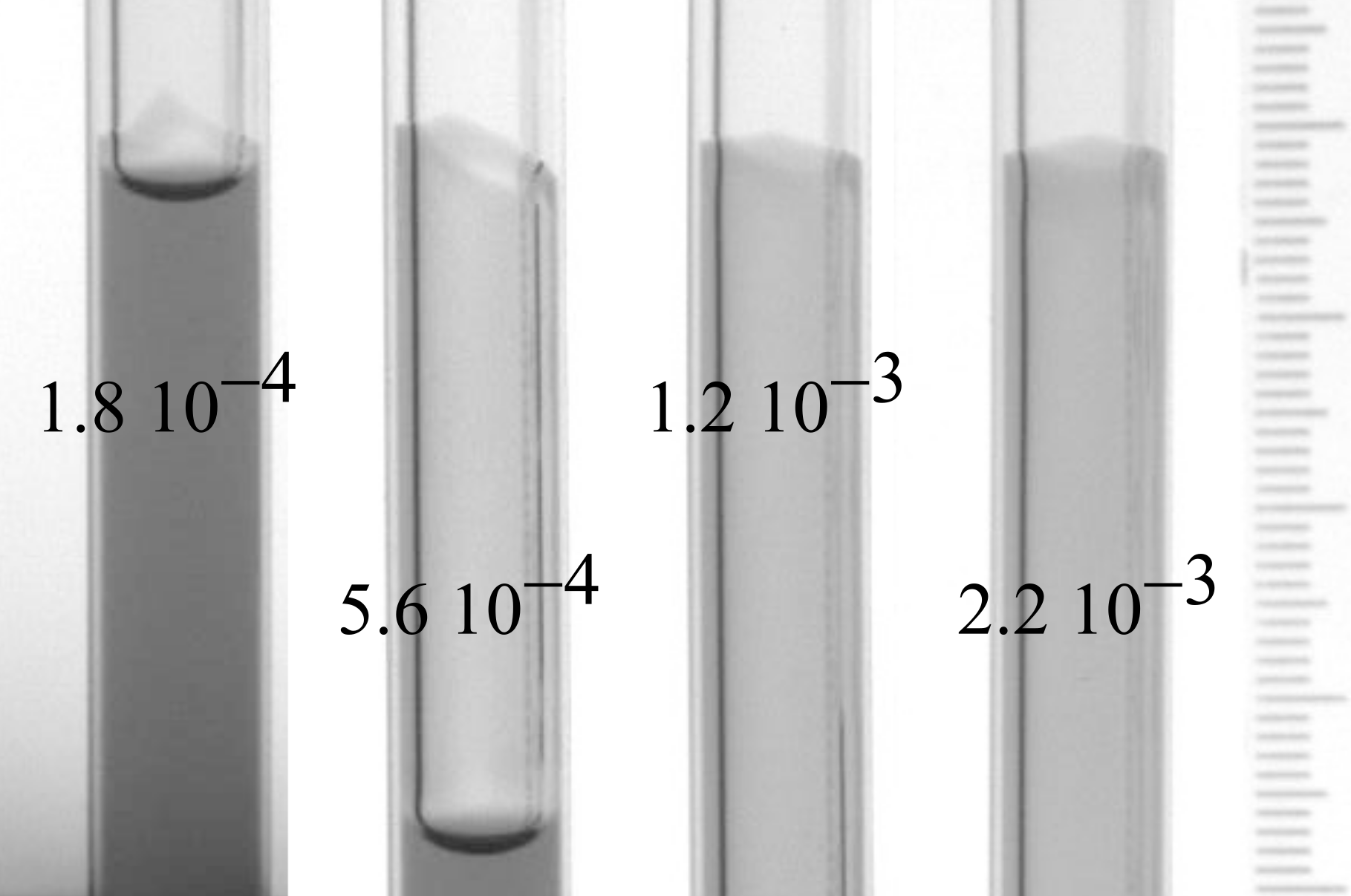}
\caption{\label{fig:curvature} Series of snapshots of dewetting films for a 30\% glycerol mixture at different meniscus capillary numbers between 1.8$\ 10^{-4}$ and 3.9$\ 10^{-3}$ showing a) the meniscus region - the dashed red lines are the predictions from the hydrodynamic model (Fig.~\ref{fig:numerics:Ca}) for each $Ca$. We find good agreement with the observed curvatures except at the lowest velocity where the film has not formed completely. At Ca=3.9$\ $10$^{-3}$, the meniscus is no longer perfectly axisymmetric due to other dynamic effects; b) the triple line region of the film (not captured at the highest velocity), showing some undulations.}
\end{figure}

In the experiments, this impact of pressure can also be observed directly on the images. Some snapshots were taken with a lower dye concentration: the film thickness can no longer be measured but the full meniscus profile is now visible. Five profiles for different velocities ranging from $Ca^*$ to slightly above $Ca^{**}$ are shown in Fig.~\ref{fig:curvature}~a) along with numerical solutions of Eq.~\ref{eq:h_steady_state} (dashed lines). Despite slight asymmetries due to some distortion in the lighting pattern, for the 3 intermediate velocities, we find that the observed meniscus shapes are consistent with the numerical predictions. The agreement between predicted and observed meniscus shapes is not as good when the film has not developed (\emph{e.g.} lowest velocity shown), or when $Ca^{**}$ has been exceeded. In this case, which is the highest velocity shown, the meniscus snapshot is no longer axisymmetric. Neither situation is accounted for in the model.

\begin{figure}
a)\includegraphics[height=5cm]{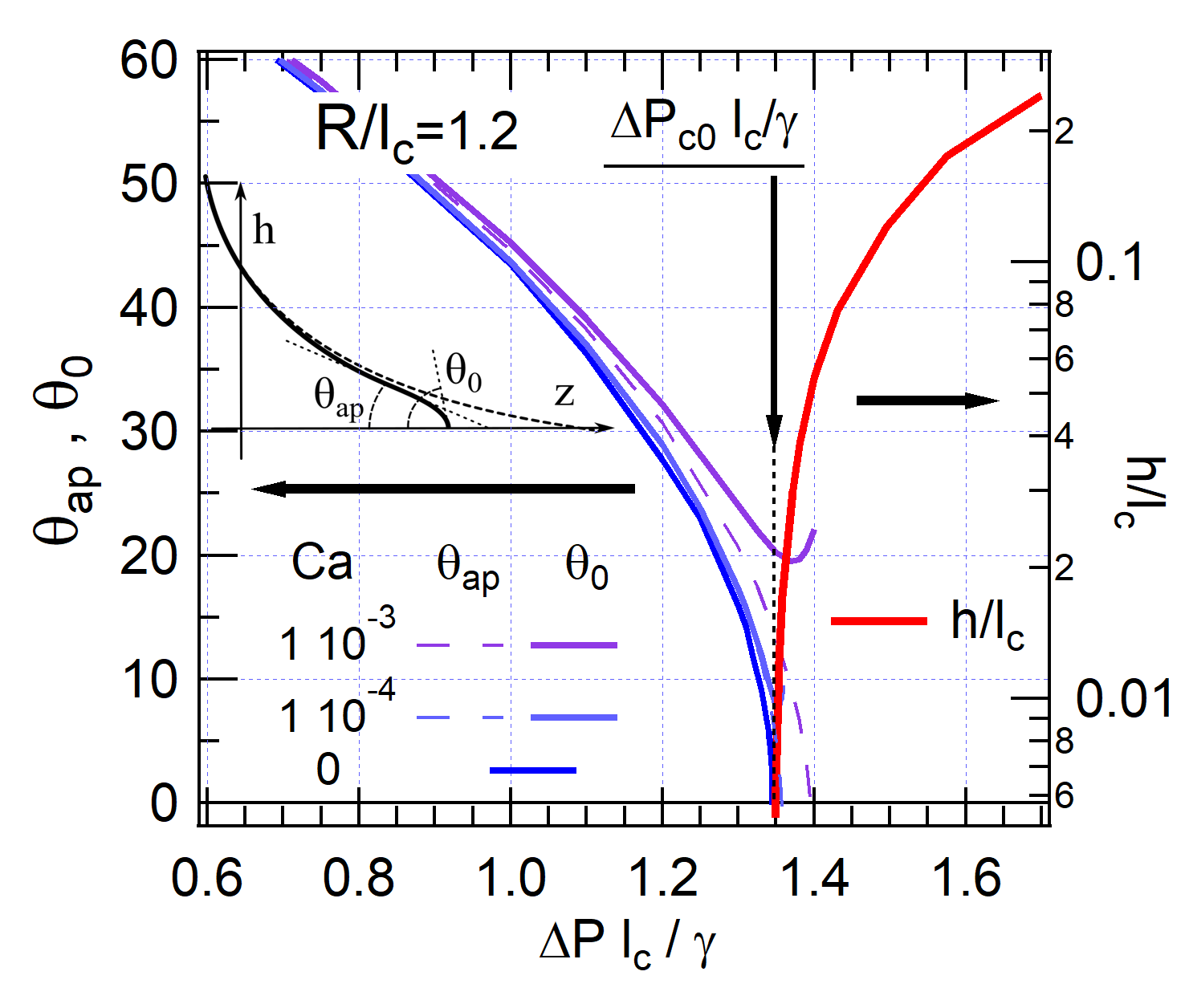}
b)\includegraphics[height=5cm]{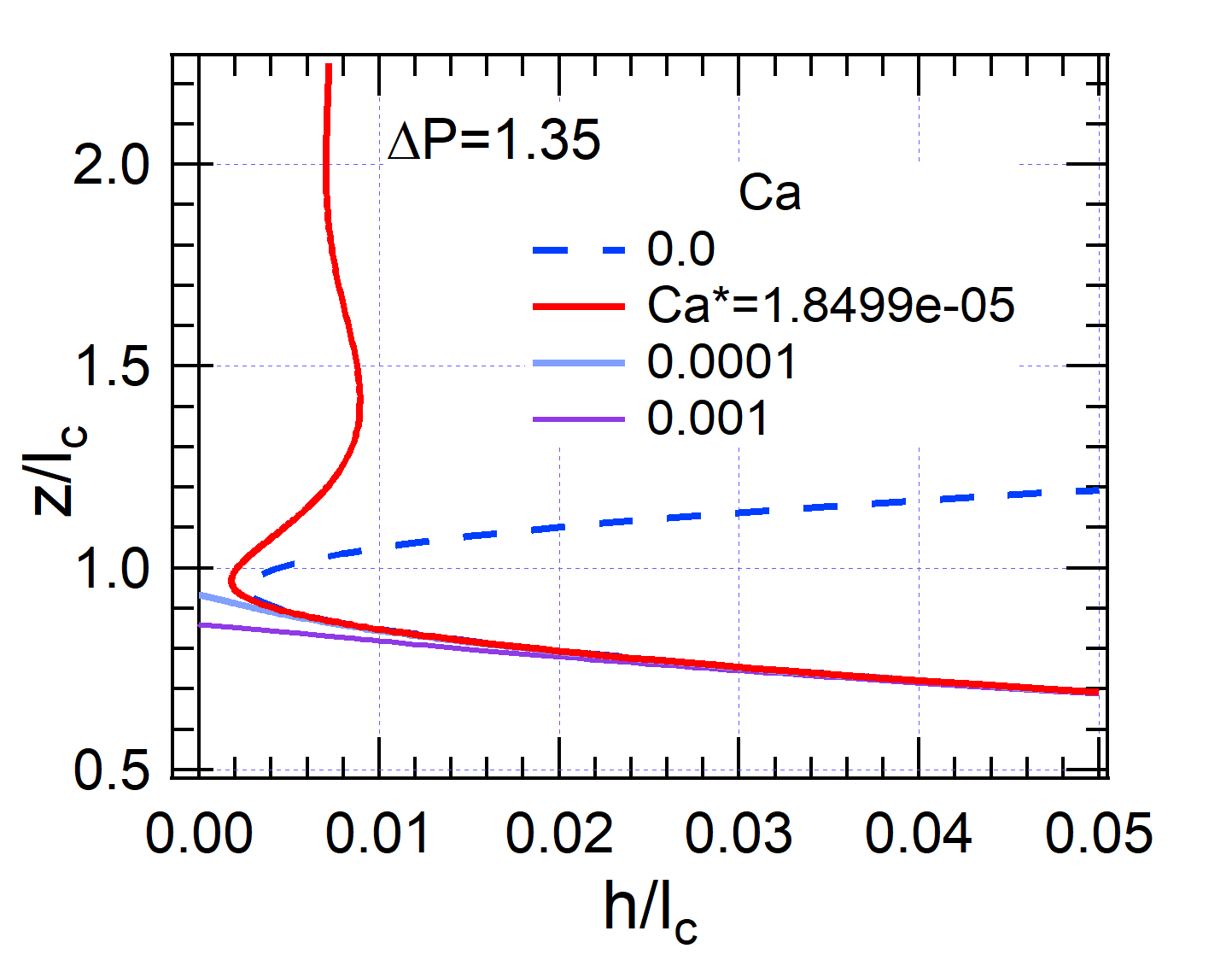}

\caption{\label{fig:DP_theta_calc} a) Calculated microscopic contact angle $\theta_0$ and apparent contact angle $\theta_{app}$ (left hand ordinate axis) as a function of pressure, for $R=1.2 l_c$ and different capillary numbers, with a slip length $\lambda/l_c$=1$\ $10$^{-4}$. The very moderate impact of dynamic effects show that in this velocity range, the solution is dominated by the pressure effect. Moreover, the contact angles drop abruptly to zero as pressure increases, which determines the critical pressure $\Delta P_c$ for the dynamic wetting transition. For $Ca=0$, the critical pressure is $\Delta P_{c0}=1.3458...$ Also shown (right hand ordinate axis) is the film thickness as a function of critical pressure for $\Delta P>\Delta P_{c0}$. Each value of thickness is associated with a different substrate velocity $Ca$, following Eq.~\ref{eq:film_thickness} (see also Fig.~\ref{fig:numerics:Ca} b); b) Meniscus profiles as a function of $Ca$ for $\Delta P>\Delta P_{c0}$. The solution at $0<Ca<Ca^*$ is bubble like and not relevant for our boundary conditions. For $Ca=Ca^*$, a thin film solution is found. For $Ca>Ca^*$, we have a meniscus with a dynamic triple line. As $Ca$ increases, the microscopic contact angle increases, \emph{i.e.} the solution is valid for a more and more hydrophobic material.}
\end{figure}

\begin{figure}
\includegraphics[height=5cm]{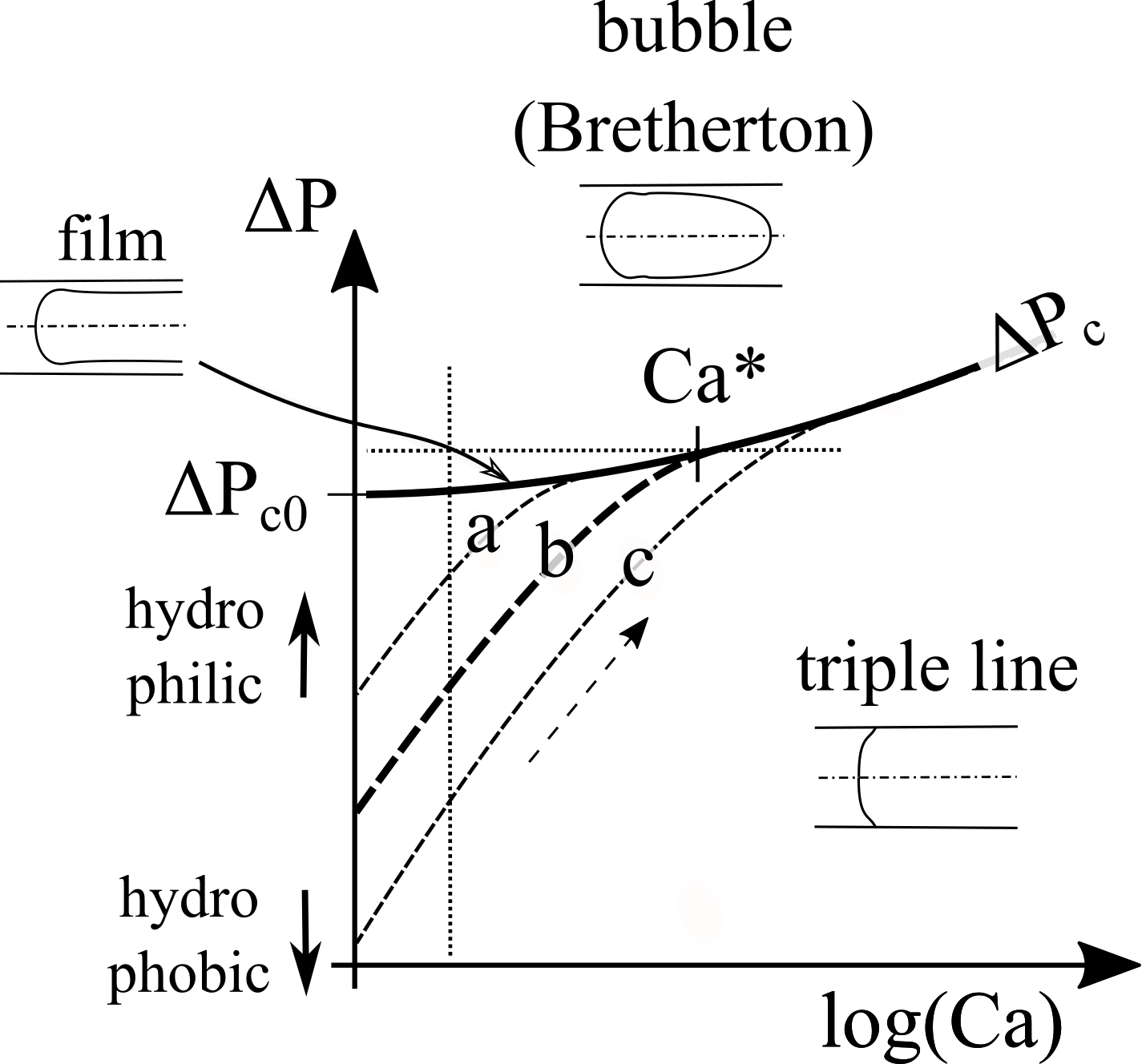}

\caption{\label{fig:DP_theta_exp} The solutions to Eq.~\ref{eq:h_steady_state} are shown in a pressure $\Delta P$ {\em vs.} velocity $Ca$ phase diagram. The high pressure bubble solutions are not allowed in our experiments due to the boundary conditions. The thin film solutions (Fig.~\ref{fig:numerics:Ca} b) form the $\Delta P_c(Ca)$ boundary (solid line). The thick dashed line (b) schematizes the family of solutions for a dynamic meniscus with a given liquid/surface couple. This line hits the $\Delta P_c(Ca)$ line at some finite velocity $Ca^*$ which also determines the minimum film thickness obtainable through Eq.~\ref{eq:film_thickness}. Applying a pressure $\Delta P>\Delta P_c(Ca^*)$ will destabilize the meniscus into a thicker film. The dashed line (a) [resp. (c)] is for a more [resp. less] hydrophilic liquid/surface couple. The more hydrophilic the surface, the lower $Ca^*$ and the smaller the minimum film thickness. The series of profiles in Fig.~\ref{fig:h_z_DP_Ca_p001} are taken along the vertical thin dashed line. The series of profiles in Fig.~\ref{fig:DP_theta_calc} are taken along the horizontal thin dashed line.}
\end{figure}

\subsection{Building up a phase diagram}\label{Sec:phase_diagram}
Intriguingly, the model seems to predict that arbitrarily thin films can be obtained: this would simply require a pressure just exceeding the threshold (Fig.~\ref{fig:numerics:Ca}b). In the experiments, in contrast, there is a lower bound to the film thickness and this limit is directly related to $Ca^*$, the velocity of the dynamic wetting transition (Fig.~\ref{fig:Ca_slow}). To clarify the role of this threshold, we now consider the stability of the dynamic meniscus, keeping in mind its dynamical properties as reflected in Fig.~\ref{fig:Ca_slow}.

Taking a constant radius $R$, we map out the space ($Ca$, $\Delta P$) of the solutions to Eq.~\ref{eq:h_steady_state}. To explore this family of solutions, we first plot the microscopic contact angle $\theta_{0}$ for static solutions ($Ca=0$) (Fig.~\ref{fig:DP_theta_calc} a - blue line). With increasing macroscopic meniscus curvature $\kappa_0$, the microscopic contact angle $\theta_0$ decreases from 90$^\circ$ at $\Delta P=0$ to 0$^\circ$ at $\Delta P=\Delta P_{c0}$. If $Ca$ is non zero, the triple line becomes dynamic, viscous bending sets in (Fig.~\ref{fig:DP_theta_calc} a - inset) and for the same $\Delta P$ the microscopic contact angle is now larger than the static microscopic contact angle. However, the difference between microscopic contact angle at finite velocity and static microscopic contact angle is rather small for our typically low values of $Ca$. It is barely noticeable for $Ca=1\ 10^{-4}$ (Fig.~\ref{fig:DP_theta_calc} - light blue line) but becomes clearer at  $Ca=1\ 10^{-3}$ (purple line). Much more significant bending would appear for 10 to 100 times larger $Ca$. To connect with the macroscopic scale problem, we also define the apparent contact angle $\theta_{ap}$ which is given by the slope of the liquid profile taken where the dynamic triple line matches the quasi-static meniscus, {\em i.e.} $h'=\tan\theta_{ap}$ at the inflexion point, where the slope $h'$ is minimum (Fig.~\ref{fig:DP_theta_calc} a - inset). With this definition, $\theta_{ap}$ is a good approximation to the experimentally measured contact angle, at the macroscopic scale such as shown in Fig.~\ref{fig:Ca_slow}.

To understand our experimental results, we must now consider the specific dynamic behaviour of the triple line and the constitutive relation between microscopic contact angle and triple line velocity. In many cases, $\theta_0$ is taken as a constant, resulting in a Cox-Voinov type relation. Experimentally, we have found a more complex evolution of $\theta_0(Ca)$ (Fig.~\ref{fig:Ca_slow}) which attests to the limited contribution of viscous dissipation at low Ca, due to low viscosity. Other dissipative mechanisms dominate in the present case, which remain elusive. However, as a result, one can neglect the small difference between $\theta_0$ and $\theta_{app}$ and the dependence shown in Fig.~\ref{fig:Ca_slow} would be a good approximation to the constitutive relation. Whatever the dissipative mechanisms, their effect can be represented by a constitutive relation $\theta_0(Ca)$.

 \added{Finally, the ($Ca$, $\Delta P$) phase diagram (Fig.~\ref{fig:DP_theta_exp}) appears as follows. We find the Bretherton type bubble at high pressures. This state is bounded at low pressures by the $\Delta P_c(Ca)$ line (cf Fig.~\ref{fig:numerics:Ca} b) which indicates the conditions under which a film forms. Below this boundary, we find the solutions with a dynamic meniscus and a triple line.} 
\added{Because for these solutions each point in the ($Ca$, $\Delta P$) phase diagram corresponds to a given contact angle, which can be calculated, specifying a constitutive relation restricts the family of solutions to a curve in this part of the phase diagram}. We have schematically shown three such $\Delta P(Ca)$ curves (dashed lines in Fig.~\ref{fig:DP_theta_exp}). Starting from zero velocity, where the static contact angle is obtained at a finite pressure (this is capillary rise), the meniscus state follows the $\Delta P(Ca)$ curve as velocity increases. For a more hydrophilic surface (curve a), the microscopic contact angle is smaller, so less viscous bending (and so a lower velocity) is required to reach equilibrium at the same pressure: the curve is shifted to the left. Conversely, for a more hydrophobic surface, the path shifts to the right (curve c). The precise path of course depends upon the details of $\theta_0(Ca)$ but the phenomenology and the physics remain the same whatever the details of the dissipation mechanisms.

Nota that the $\Delta P_c(Ca)$ line, which marks the appearance of the thin film, does not depend upon the constitutive relation of the triple line since the liquid surface does not meet the solid. However, the dynamic meniscus becomes unstable and a spreading thin film appears when, for a given liquid surface couple, the $\Delta P(Ca)$ curve meets the $\Delta P_c(Ca)$ line. This will occur at the finite velocity $Ca^*$ and result in the corresponding finite film thickness. Thus, in the pressure range $\Delta P_{c0}<\Delta P < \Delta P_{c}(Ca^{*})$, the dynamic meniscus solution is stable. The corresponding range of film thickness is inaccessible. As a further example, a series of solutions calculated at fixed pressure slightly above $\Delta P_{c0}$ for different values of $Ca$ is shown in Fig.~\ref{fig:DP_theta_calc}b. The static solution $Ca=0$ is bubble-like and not relevant for our boundary conditions. A finite velocity is required with enough viscous bending to meet the surface. In fact, at the threshold velocity $Ca^*$, a zero flux film forms. For $Ca>Ca^*$, the dynamic meniscus has a triple line.

Thus, although the relation between pressure and film thickness does not depend upon the liquid/surface properties, the threshold value $Ca^*$ does ~\cite{hocking_meniscus_2001,eggers_contact_2005}. In the experiments, preliminary tests were carried out with the same liquids and in tubes identical in geometry but made of polydimethylsiloxane (PDMS). The results demonstrate a similar phenomenology but with $Ca^*=3.4\ 10^{-4}$, a ten fold increase which is easily rationalised by the more hydrophobic nature of PDMS ($\theta_e=80^\circ$). However, because these tubes are not optically clear, more advanced evaluations could not be carried out.

\section{Discussion}
We have given a detailed account of the impact of pressure on the stability of the dynamic meniscus in a tube. At low pressures, the meniscus curvature is low enough and a suitable velocity can be found to ensure that local contact angle dynamics and equilibrium equations are obeyed simultaneously.
For larger pressures, $Ca>Ca^*$ and the triple line is unstable so that when pressure is applied, a thin film forms almost instantaneously. Its thickness depends upon the meniscus curvature as shown in Fig.~\ref{fig:numerics:Ca} b. Pressure is a control parameter from which film thickness can be adjusted (Fig.~\ref{fig:exp:profiles:Ca}). A significant range of steady state film thicknesses can be selected, in contrast to previous results~\cite{hocking_meniscus_2001,snoeijer_avoided_2006} where a unique film thickness was selected, dependent on the uniquely valued $Ca$ at the forced wetting transition. The lowest accessible thickness (and thus the critical capillary number $Ca^*$) depends upon the liquid surface properties.

Once formed, the film adopts a given thickness but it can adapt to a range of meniscus velocities, as shown in Fig.~\ref{fig:instationary}. Indeed, within the lubrication approximation, Wilson and Jones~\cite{wilson_entry_1983} have shown how, due to the gravity driven downward flow, the film matches to the meniscus through two oscillatory solutions with one phase-like free parameter. With this additional variable, different meniscus velocities can be accommodated by the same film thickness. Interestingly, the oscillatory nature of the predicted profiles is directly reflected in the dimple we observe at the junction between film and meniscus (Fig.~\ref{fig:exp:profiles:Ca} c). In contrast, in the Landau Levich solution, there is no gravity driven flow, and there is a single solution of exponential type. In fact, in our case, when the velocity is too large, the downward, gravity driven flux becomes too small and the steady state solution switches to a  Landau-Levich, single exponential, type. This transition marks the upper limit to the regime of interest here and the threshold velocity is the upper bound $Ca^{**}$ mentioned earlier.

We now turn to the relation between the film and its own triple line, at the top. Because the steady state differential equation is of third order, while there are two oscillatory solutions at the bottom, there is a single exponential-like solution at the top to match the film to the triple line. It is this structure of the solution which explains why, given the local dynamics of dewetting $\theta_0(Ca)$, the drainage of a film fed by a dewetting front at its top leads to a unique thickness value and unique receding triple line velocity, as discussed earlier~\cite{hocking_meniscus_2001,snoeijer_avoided_2006}.

Surprisingly, in our case, we find that the triple line at the top of the film can accommodate \emph{a range} of dewetting velocities which are actually selected by the applied pressure as the film emerges from the unstable meniscus. We do  not have a definite explanation for this observation, but we can point out a few features of the triple line of the film which do not belong to the standard description~\cite{hocking_meniscus_2001,snoeijer_avoided_2006}. First we observe a (rather flat) steady state rim which does not belong to the standard model. Also, the triple line is usually not straight but has an undulating profile and may even sometimes exhibit angular features (Fig.~\ref{fig:curvature} b). Modulations of the shape of the triple line and especially such angular features are indeed known to regulate dewetting speeds above the entrainment threshold~\cite{blake_maximum_1979,podgorski_corners_2001, snoeijer07_cornered} while line front modulations accelerate the relaxation dynamics~\cite{snoeijer_relaxation_2007}. Therefore, although the exact mechanism is not clear, we suspect that it is such additional dynamic phenomena connected to modulation of the contact line profile which stabilize the film thickness.

\section{Conclusion}
In conclusion, we have investigated the dynamic wetting transition for partially wetting liquids inside tubes. With this geometry, we can reach relatively high velocities and thus use low viscosity fluids. In addition, using absorption imaging, we can accurately follow the thin film geometry and dynamics at high speed. Our observations of the dynamic wetting transition and the geometrical and dynamic characteristics of the liquid film mostly conform to standard models. However, in place of the unique film solution predicted and reported in the literature, we find that we can select the film thickness within a well identified range. The selection operates immediately upon destabilization of the \added{dynamic meniscus}. It is mediated by the pressure induced meniscus curvature, thus providing a direct experimental demonstration of the effect of macroscopic geometry on the stability of a dynamic meniscus. The critical capillary number $Ca^*$ above which a film forms (and thus the lowest accessible thickness) depends upon the liquid/surface properties. The film thickness is directly related to the velocity of the dewetting triple line at the top of the film as expected from the Derjaguin relation. However, the very fact that the film thickness is not unique, but can be tuned over almost a decade, does not conform to the standard model. This observation implies a more complex dynamics for the triple line which may arise from its undulating morphology at the macroscopic scale, a feature which stands at variance with the standard assumption of a straight front. This high velocity dynamics of the triple line certainly deserves further studies.



\begin{acknowledgments}
We thank Ludovic Olanier for help with the experimental set-up and Arnaud Antkowiak, Terry Blake, Marc Fermigier, Laurent Limat, David Qu\'{e}r\'{e}, Etienne Reyssat and Jacco Snoeijer for fruitful discussions and suggestions.
\end{acknowledgments}

\bibliography{my_references_PH_160531}
\bibliographystyle{unsrt}

\appendix
\section{Normalized equations}\label{app:norm_eq}
All spatial dimensions are normalized by $l_c=(\gamma/{\rho g})^{1/2}$ including film thickness $h=l_c\tilde h$, vertical coordinate $z=l_c\tilde z$ and tube radii $R$. Pressures $\Delta P$ are normalized by $\gamma/l_c$ and thus equal to normalized curvatures $\tilde \kappa = l_c \kappa$. Then Eqs.~\ref{eq:h_steady_state} and \ref{eq:kappa_steady_state} become
\begin{equation}\label{eq:h_steady_state_norm}\frac{{\tilde h}^3}{3} \left( \partial_{\tilde z} \tilde\kappa  + 1 \right) - Ca\ {\tilde h} =  cte 
\end{equation}
and
\begin{equation}\label{eq:kappa_steady_state_norm}\tilde\kappa = -\frac{\partial^2_{{{\tilde z}}{{\tilde z}}} {\tilde h}}{\left(1+(\partial_{\tilde z} {\tilde h})^2\right)^{3/2}}+\frac{1}{\left(\sqrt{Bo}-{\tilde h}\right)\sqrt{1+(\partial_{\tilde z} {\tilde h})^2}}\end{equation}
with $Ca=\eta v/\gamma$ and $Bo=(R/l_c)^2$.

\section{Impact of tube radius}
\begin{figure}
a)\includegraphics[height=6cm]{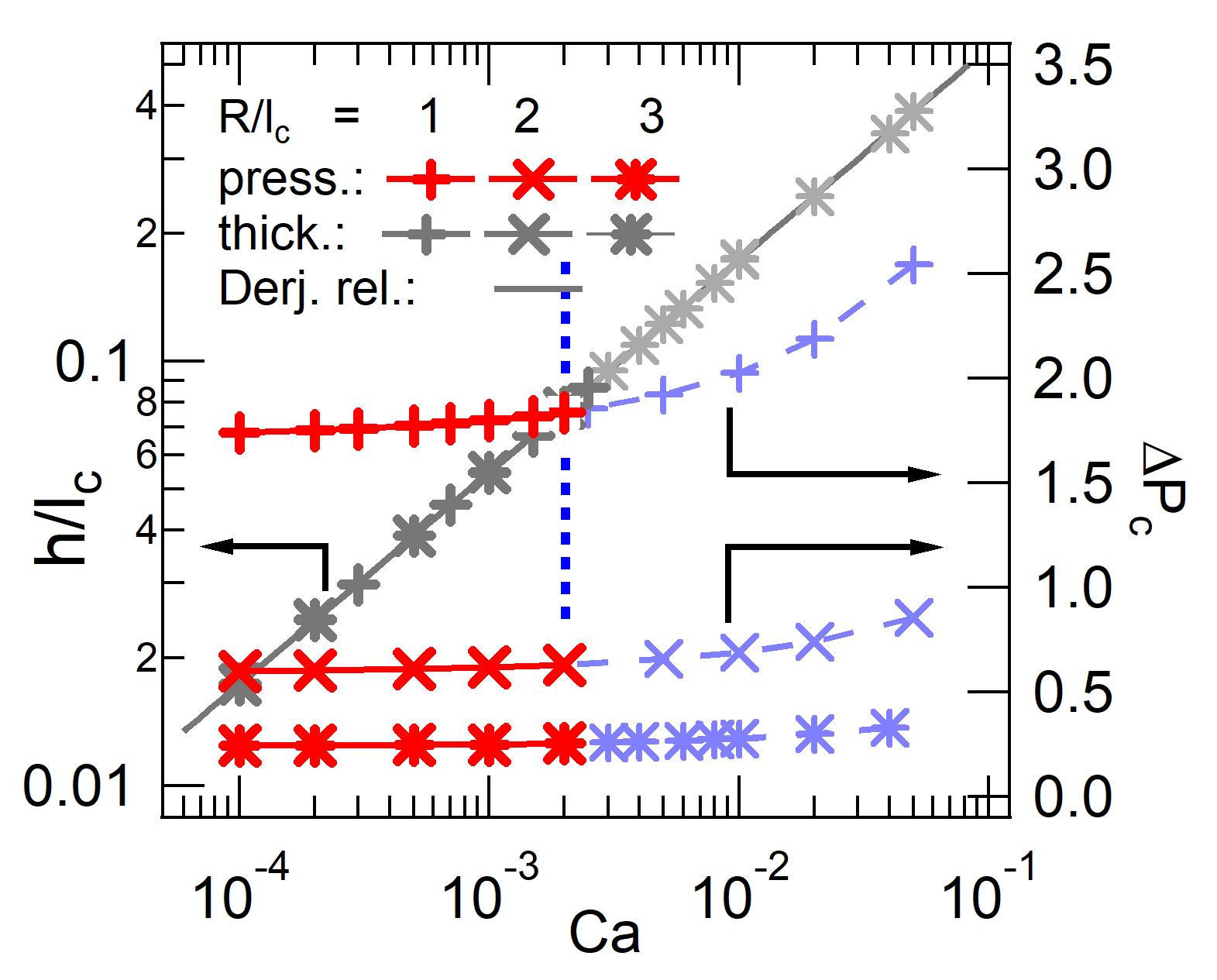} 
b)\includegraphics[height=6cm]{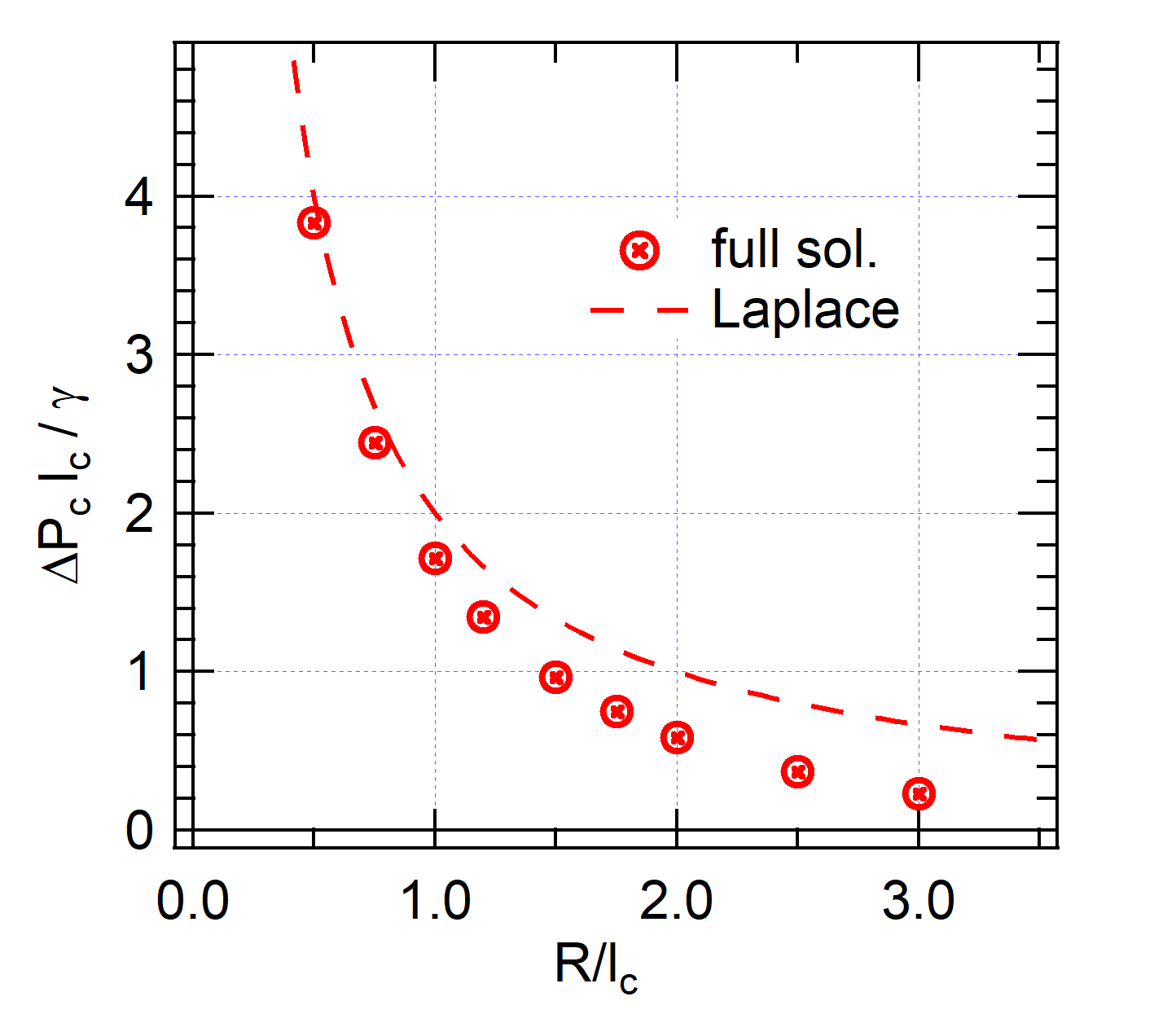}
\caption{\label{fig:DP_vs_R}a) critical pressures
and corresponding film thicknesses for different meniscus velocities $Ca$ (cf Fig.\ref{fig:numerics:Ca} b) -- impact of the tube radius (Bond number); b) critical pressure at zero velocity as a function of tube radius.}
\end{figure}

 To better understand the role of gravity, the dynamic meniscus calculations were performed for different tube radii (Fig.~\ref{fig:DP_vs_R} a). The film thickness is independent of the tube radius, as expected from Eq.~\ref{eq:film_thickness} but as the tube radius increases, the critical pressure $\Delta P_c$ decreases and the pressure variations become weaker. The variations of the critical pressure at zero velocity $\Delta P_{c0}$ with tube radius is shown in Fig.~\ref{fig:DP_vs_R} b. For $R<l_c$, $\Delta P_{c0}\simeq\gamma/(2R)$, which is the Laplace pressure. Owing to gravity, $\Delta P_{c0}$ drops faster towards 0 when $R > l_c$. At larger radii, the flat plate limit is recovered where solutions are pressure independent because they are dominated by the intrinsic, gravity driven, curvature of the profile near the tube surface.

\end{document}